\documentclass{aa}  

\usepackage{graphicx}
\usepackage{txfonts}
\usepackage{lipsum}
\usepackage{subcaption}         
\usepackage{lscape}   
\usepackage{lineno}
\usepackage{placeins}         
\usepackage{hyperref}

\usepackage{lipsum}
\usepackage{xspace}
\usepackage{amsmath}
\usepackage{soul}
\usepackage{supertabular}
\usepackage{booktabs}

\linenumbers
\renewcommand\makeLineNumber{}

\newcommand{\halpha}{H$\alpha$\xspace}

\newcommand{\vhe}{$V_{\text{He}}$\xspace}

\newcommand{\kms}{{\rm ~km s$^{-1}$}\xspace}
\newcommand{\msun}{{\rm M$_{\odot}$}\xspace}

\usepackage{xcolor}
\usepackage{xspace}

\begin{document}
\title{NLTE Spectral Modelling of the Nearby Stripped-Envelope Supernova 2024ehs}

\author{
W. Sand Hellman\inst{1}
\and
S. J. Brennan\inst{2,1}
\and 
S. Bartmentloo\inst{1}
\and
C. Fremling\inst{3,4}
\and
A. Gangopadhyay\inst{1}
\and
A. Jerkstrand\inst{1}
\and
R. Lunnan\inst{1}
\and
J. Purdum\inst{3}
\and
S. Schulze\inst{5}
\and
J. Sollerman\inst{1}
}

\institute{
The Oskar Klein Centre, Department of Astronomy, Stockholm University, AlbaNova, SE-10691 Stockholm, Sweden
\and
Max Planck Institute for Extraterrestrial Physics, Max-Planck-Gesellschaft, Giessenbachstraße 1, Garching, 85748, Germany.
\and
Caltech Optical Observatories, California Institute of Technology, Pasadena, CA 91125, USA
\and
Division of Physics, Mathematics and Astronomy, California Institute of Technology, Pasadena, CA 91125, USA
\and
Department of Particle Physics and Astrophysics, Weizmann Institute of Science, 234 Herzl St, 76100 Rehovot, Israel
} 

\abstract{
We present a detailed study of the Type IIb supernova 2024ehs, discovered in March 2024 in the nearby galaxy NGC 3443 at a distance of $23.8\pm 0.9$ Mpc. Using photometric and spectroscopic observations spanning 10 months, we analyse its light curve, spectral evolution, and physical properties with the \texttt{SUMO} radiative transfer code. SN 2024ehs exhibits a narrow light-curve peak, rapid decline, and weak helium lines, distinguishing it from typical Type IIb supernovae. Comparisons with other objects, including SNe 1993J and 2020acat, and modelling of nebular spectra suggest a low ejecta mass, high velocities ($\sim$20,000 \kms), and a $^{56}$Ni mass just below $\sim0.1 \, \mathrm{M}_{\odot}$. Furthermore, the nebular spectral models indicate a progenitor with a helium core mass of $\sim 6 \, \mathrm{M}_{\odot}$, consistent with a zero-age main-sequence mass of $\sim 23\, \mathrm{M}_{\odot}$. The composition of spectra is explored through photospheric modelling, finding a link between expansion velocity and the relative strength of different element lines. This work discusses further the diversity of stripped-envelope supernovae and the role of binary interactions for their progenitors, and demonstrates the need for further modelling to refine $^{56}$Ni mass estimates and to understand the physical mechanisms driving their evolution.
}

\keywords{Core-collapse supernovae --- Stripped-envelope supernovae --- Radiative transfer}

\maketitle

\section{Introduction} \label{sec:intro}

The classification of supernovae (SNe) is originally based on spectral features \citep{Baade1934,Minkowski1941}. The initial divide between Type I and II was established based on  the absence or presence of hydrogen lines in the optical spectra \citep{Minkowski1941}. This classification scheme has been expanded as specific sub-types have emerged based on the shape and presence of other spectral lines \citep{Filippenko1997, Gal-Yam2017}. 


Type IIb SNe, first discovered with the object SN 1987K \citep{Filippenko1988}, are a subclass of supernovae that initially display strong hydrogen lines typical of a Type II, but morph with time into something closer to a Type Ib with dominant helium lines \citep{BranchWheeler}. The progenitor's history likely involves stages of enhanced mass loss, resulting in the terminal SN explosion displaying weak-to-intermediate emission from hydrogen -- reflecting the low mass of the remaining hydrogen envelope. The classification of Type IIb was adopted to reflect this middle ground between types \citep{Gilkis2022}. The loss of the outer hydrogen envelope  can occur in multiple ways, including strong winds and mass transfer between binary companions \citep{Sana2012,Marchant2024}. Mass loss due to strong winds can occur in massive stars preferably with high metallicity \citep{Puls2008,Krticka2014}, so-called Wolf-Rayet (WR) stars \citep{ Shenar2024}. These stars are typically depleted of their hydrogen envelopes, and in some cases also lack helium, and would explode as stripped-envelope SNe (SESNe). 

The progenitor mass (before mass loss) of a typical SESN is much lower than the mass required to produce a WR star at solar metallicity. The observed abundance of SESNe is also too high to be explained solely by WR stars \citep{Smith2011}. Massive stars in binaries, on the other hand, are much less constrained, and massive stars preferably form in binary systems \citep{sanaEvans2011}. Up to 50\% of all stars in binaries experiencing mass transfer between the companions \citep{Sana2012}.

There are three SESNe with confirmed detections of binary companions: SN 1993J \citep{Maund2004, Fox2014}, SN 2001ig \citep{Ryder2018} and SN 2011dh \citep{Maund2015}. All of these are Type IIb, confirming that binary mass transfer is at least one of the formation tracks for these objects. The mass of the remaining hydrogen shell is expected to be below $\sim 0.5~\mathrm{M}_{\odot}$ \citep{Yoon2017,Gilkis2022}, but the exact limit between Type IIb and Ib is still contested. Some evidence suggests that mixing affects the amount of hydrogen that can stay ``hidden'' in an object without significant signs in the spectra \citep[e.g; ][]{Chen2018,Gangopadhyay2023}. \cite{Dessart2011} found, through non local thermodynamic equilibrium (NLTE) simulations of light curves and spectra, that a hydrogen mass as low as $\sim10^{-3}~M_{\odot}$ is enough to produce the spectrum of a Type IIb. \cite{Hachinger2012} found a lowest limit $\sim0.033~\mathrm{M}_{\odot}$ of hydrogen required to transition between Types IIb to Ib. \cite{Gilkis2022} have looked at simulations compared to observational data and agree with a lowest hydrogen mass limit of $\sim 0.033~\mathrm{M}_{\odot}$ for Type IIb. 

\cite{Chevalier2010} suggest a subdivision within Type IIb, between what they call extended and compact objects. Compact objects do not display an extended shock cooling phase in their light curve and typically lack hydrogen in the nebular phase. They find that the crucial difference between extended and compact objects is the mass of the hydrogen envelope, where extended Type IIb may have an envelope mass above $\sim0.1~\mathrm{M}_{\odot}$. They also suggest that compact Type IIb are likelier candidates for WR progenitors. 

This paper presents data on the stripped-envelope SN 2024ehs, discovered in March 2024 and monitored until November of the same year. Through spectral comparisons to other Type IIb as well as spectral modelling of the late photospheric and nebular phase spectra we attempt to further our understanding of what the spectral behaviour tells us about the progenitor. 

\section{Discovery and Follow-Up Observations}\label{sec:discovery}

\subsection{Discovery of SN 2024ehs and Distance to NGC 3443}\label{ssec:distance}

SN 2024ehs was initially reported by the Asteroid Terrestrial-impact Last Alert System (ATLAS) collaboration \citep{Tonry2024} and noted to be a rapidly brightening candidate SN in NGC 3443 \citep{Smith2024}. SN 2024ehs is a very nearby source, located in the Intermediate Spiral Galaxy NGC 3443 in the constellation Leo. 

While the close proximity of the object has allowed for observations up to almost 10 months after discovery despite its relatively low intrinsic luminosity, the relative uncertainty in the distance measurements is significant, and this is expected to affect all calibrations of the data. Throughout this work, a distance of $23.8\pm 0.9$ Mpc has been assumed. This is the mean of multiple Tully-Fisher measurements of the distance to the host galaxy reported in the {NASA/IPAC Extragalactic Database} (\textsc{NED}\footnote{The NASA/IPAC Extragalactic Database (NED) is operated by the Jet Propulsion Laboratory, California Institute of Technology, under contract with the National Aeronautics and Space Administration.}), though it is crucial to note that distances reported in \textsc{NED} range between 19.3 -- 26.9 Mpc, a significant uncertainty.

\subsection{Photometry}\label{ssec:photometry}

\begin{figure}[!ht]
    \includegraphics[width=1\linewidth]{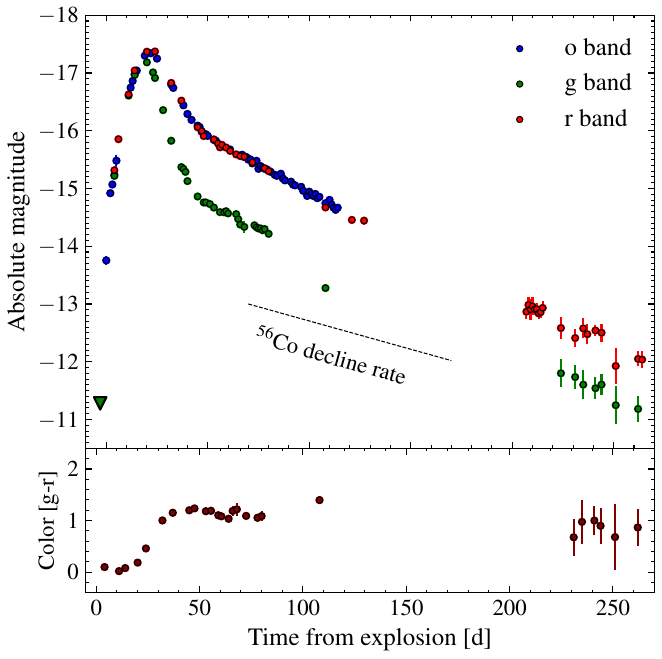}
    \caption{Light curves of SN 2024ehs, in $o$, $g$, and $r$ bands. The data points have been re-binned into bins of $\sim 1$ day. The decay rate for $^{56}$Co is 0.98 mag per 100 days and is provided as a comparison to the tail of the light curves.}
    \label{fig:lc}
\end{figure}

Science-ready images obtained as part of the Zwicky Transient Facility (ZTF) survey \citep{Bellm2019, Graham2019, Dekany2020, Masci2019} were acquired in $gr$-bands from the NASA/IPAC Infrared Science Archive (IPAC\footnote{\href{https://irsa.ipac.caltech.edu/applications/ztf/}{https://irsa.ipac.caltech.edu/applications/ztf/}}) service. Template-subtracted photometry was performed using the \texttt{AutoPhoT} pipeline \citep{Brennan2022d} using ZTF deep template images. Photometry from the Asteroid Terrestrial-impact Last Alert System (ATLAS) forced-photometry server\footnote{\href{https://fallingstar-data.com/forcedphot/}{https://fallingstar-data.com/forcedphot/}} \citep{Tonry2018,Smith2020,Shingles2021a} was obtained in the $o$ band. We computed the weighted average of the fluxes of observations on a nightly cadence. A quality cut at 5$\sigma$ was applied to the resulting nightly fluxes for both filters, after which the measurements were converted to the AB magnitude system.

Fig.~\ref{fig:lc} shows the light curve of SN 2024ehs in $r$, $o$ and $g$ band. The data points for the individual bands have been re-binned into bins of 1 d. SN 2024ehs was first discovered on March 15, MJD 60383.4 (with the last non-detection on MJD 60380.4 at g > 20.6 mag), and peaked 19, 22.5 and 23.5 days later in $g$, $o$ and $r$ band, respectively. The absolute brightest magnitude measured was $-17.37$ at peak in the $r$ band. Past $\sim 55$ days, the light curve is fading linearly, resembling a radioactive nickel tail in all three bands. The expected radioactive decay of $^{56}$Co $\rightarrow ^{56}$Fe with full $\gamma$-trapping can be estimated as 0.98~mag/100d \citep{patat1994}. From $\sim$55 d after peak in all bands, the tail-phase slope of SN 2024ehs is $\sim$1.82~mag/100d, significantly steeper than the $^{56}$Co decay. This is consistent with the idea that SESNe do not achieve full $\gamma$-trapping at any point in their evolution \citep[e.g.][]{jerkstrand2025corecollapsesupernovae}, as well as earlier calculations finding the decline rate of Type IIb SNe to be approximately double that of $^{56}$Co \citep{Ergon2015}. The $g-r$ colour increases slightly around peak, as the g band drops off quicker, but remains stable when both bands reach the nickel tail, until last detection.

\begin{figure*}[!ht]
    \centering
    \includegraphics[scale=0.6]{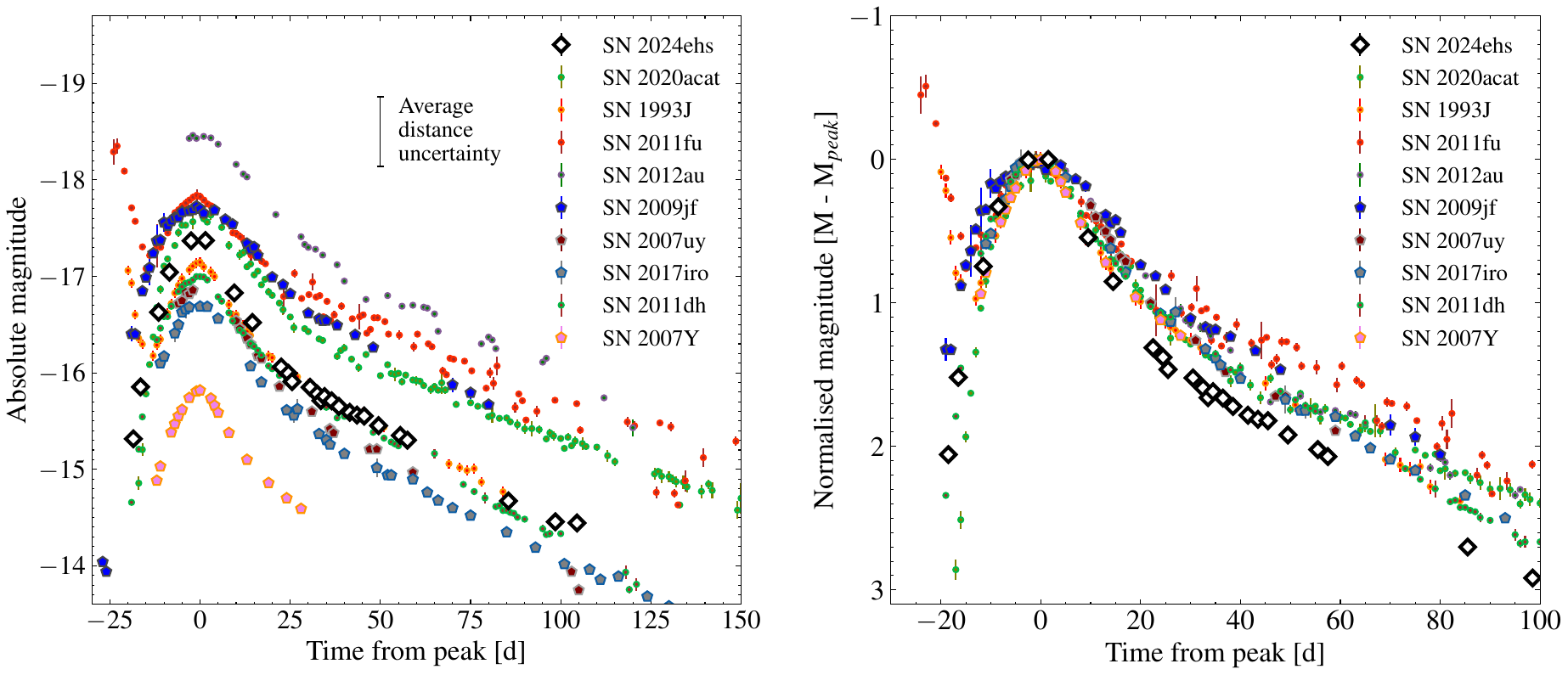}
    \caption{Comparison of the light curve absolute magnitude (left), and shape (right) between SN 2024ehs and those of SNe 1993J \citep{Filippenko1993}, 2020acat \citep{ergon2023}, 2011dh \citep{sahu2013} 2011fu \citep{Kumar2013,MoralesGaroffolo2015}, 2007Y \citep{Stritzinger2009}, 2009jf \citep{Valenti2011}, 2007uy \citep{Roy2013}, 2011iro \citep{Kumar2022} and 2012au \citep{Pandey2021}. For the comparison objects, dots represent Type IIb and pentagons represent Type Ib. All light curves are taken in R band, except SN 2024ehs and SN 2007Y, both in r band. The isolated error bar is to visualise the spread in absolute magnitude between the different distance estimates of NGC 3443 reported in NED.}
    \label{fig:comp_lc}
\end{figure*}

Fig.~\ref{fig:comp_lc} shows a comparison between the $r$-band light curve of SN 2024ehs and similar objects including the Type IIb SNe 1993J \citep{Filippenko1993}, 2020acat \citep{ergon2023}, 2011dh \citep{sahu2013} and 2011fu \citep{Kumar2013,MoralesGaroffolo2015} as well as Type Ib SNe 2007Y \citep{Stritzinger2009}, 2009jf \citep{Valenti2011}, 2007uy \citep{Roy2013}, 2011iro \citep{Kumar2022} and 2012au \citep{Pandey2021}. SN 2007Y is also given in $r$-band (calibrated to AB magnitudes), while the rest of the comparison objects are in $R$ band. While SN 2024ehs has a somewhat typical peak magnitude and a normal (though on the narrower side) light-curve peak, it displays a relatively steep decline in its tail phase compared to other objects. This may imply that the ejecta mass is lower in SN 2024ehs \citep{Arnett1982}, and that the $\gamma$-trapping is particularly ineffective.


We also do not see an initial cooling phase from shock breakout in the light curve, as can be seen in objects like SN 1993J. Our data extends early enough before peak that it is reasonable to assume that this is not because the shock-cooling phase was missed by detections. It is expected that the duration of the shock breakout is connected to the extension of the progenitor, which would mean that SN 2024ehs has a relatively compact progenitor, with no surrounding shell structure \citep{falk1977}. 

\subsection{Spectra}\label{ssec:spectra}

\begin{figure*}[!ht]
    \centering
    \includegraphics[width=\textwidth, scale=0.5]{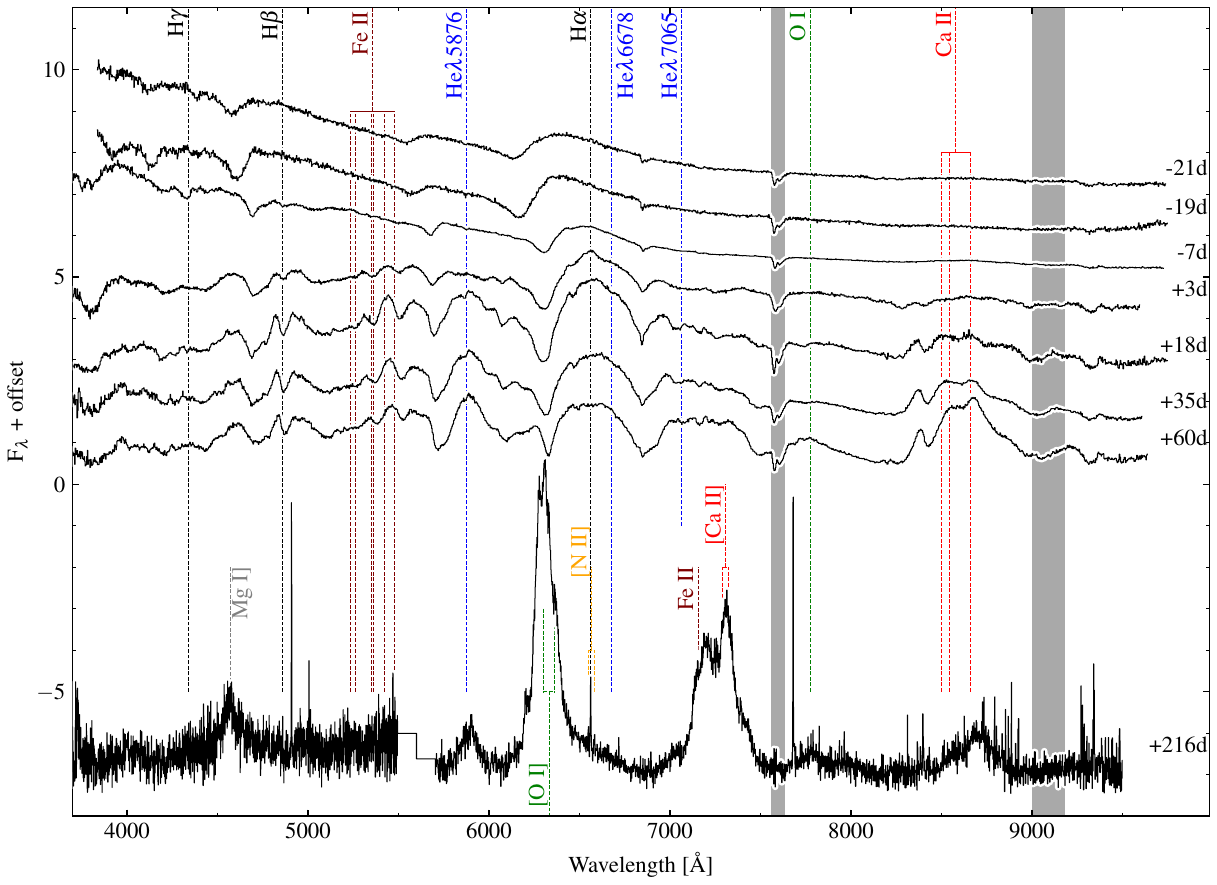}
    \caption{The evolution of the spectra from SN 2024ehs. Normalised and with added offsets for comparison. We mark atmospheric telluric absorption as grey bands and the rest wavelenghts of some significant lines. Time is relative to date of peak in the $r$-band.}
    \label{fig:spec_ev}
\end{figure*}

Our spectral sequence\footnote{Spectra for SN 2024ehs are available at  https://www.wiserep.org/object/24939} is mainly taken using the Alhambra Faint Object Spectrograph and Camera (ALFOSC) instrument on the Nordic Optical Telescope (NOT). Spectra were reduced and calibrated in a standard manner using a custom \texttt{PypeIt} \citep{pypeit:zenodo,pypeit:joss_arXiv,pypeit:joss_pub} code\footnote{\href{https://gitlab.com/steveschulze/pypeit_alfosc_env}{https://gitlab.com/steveschulze/pypeit\_alfosc\_env}} with observations coordinated using the \texttt{FRITZ} data platform \citep{Walt2019,Coughlin2023}. Fig.~\ref{fig:spec_ev} shows the evolution of the optical spectra of SN 2024ehs, with days given in reference to the $r$-band peak. Each individual spectrum has been scaled to the $r$-band photometry. The most notable feature is the apparent absence of the He~I $\lambda$6678 emission profile. This line characteristically shows up as an absorption dividing the H${\alpha}$ emission during the post-peak phase \citep{BranchWheeler}. Instead, the H${\alpha}$ line in the spectra of SN 2024ehs develops into a boxy, top-hat line shape with only minor perturbations that could be interpreted as absorption features. In the post-peak evolution, multiple lines around 5500 ~$\AA$ develop, typically a sign of Fe II, as well as the Ca II NIR triplet around 8500 ~$\AA$. The initial velocities vary slightly between lines, with the hydrogen absorption features measuring around $\sim 20000$ \kms\ in the first spectrum. This is further explored in Sect.~\ref{ssec:vels}.

Our nebular spectrum acquired 7 months post peak was taken using the \textsc{DBSP} spectrograph on the Palomar 200-inch telescope. Data were reduced using the \textsc{DBSP\_DRP} \cite{Mandigo-Stoba2022} package, and have again been scaled to the $r$-band photometry. Dominant features in the nebular spectrum include the [O I] $\lambda\lambda$6300,6364 and [Ca II] $\lambda\lambda$7291,7323 doublets. Also visible, though weaker, are the Mg I] $\lambda$4571 and a persistent bump from [Ca II] $\lambda$8662, as well as the Na I D doublet at $\lambda\lambda$5896,5890. The [Ca II] doublet is also likely blended with the [Fe II] $\lambda$7155 line, which seems unusually strong in this object. The [Ni II] $\lambda\lambda$6548,6583 doublet is present but very weak. The nebular spectrum is used for progenitor mass estimation through spectral modelling, discussed further in Sect.~\ref{ssec:sumo}.

\begin{figure}[!ht]
    \centering
    \includegraphics[width=1\linewidth]{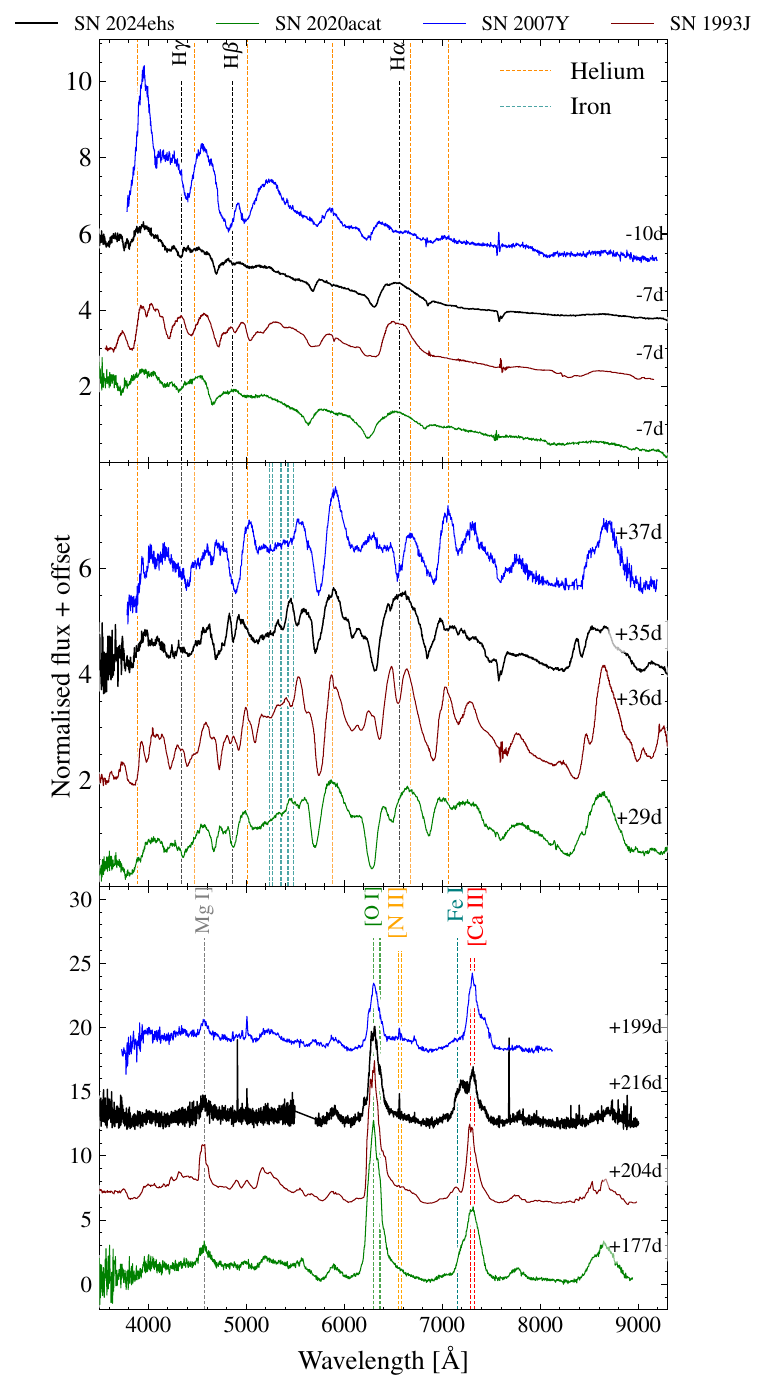}
    \caption{Spectra from three different phases in the evolution of SN 2024ehs; rising (top), $\sim$a month post-peak (middle) and nebular (bottom), as compared to three objects in similar phases; Type Ib/Ic 2007Y \citep[in blue, ][]{Stritzinger2009} and Type IIb SNe 1993J \citep[in maroon, ][]{Matheson2000} and 2020acat \citep[in green, ][]{Medler2022acat}. Some significant emission lines for each phase are marked at rest wavelengths.}
    \label{fig:spec_comp}.
\end{figure}

Spectra from SN 2024ehs taken in three different phases of its light curve evolution are compared to spectra of similar objects in the same phase in Fig. \ref{fig:spec_comp}. SN 2007Y \citep{Stritzinger2009} is chosen due to its similarity with SN 2024ehs's light curve evolution, and SN 1993J \citep{Filippenko1993} as an example of a typical Type IIb. SN 2020acat \citep{Medler2022acat} is used in this comparison because it demonstrates a similar weakness in the He~I $\lambda$6678 absorption as SN 2024ehs. Through this comparison it is seen that even though SN 2024ehs matches very well with Type Ib objects in its light curve, there is a clear hydrogen abundance in its spectra compared to SN 2007Y, strongly favouring a Type IIb classification. 

\cite{Medler2022acat} noted that SN 2020acat has an unusually weak He~I $\lambda$6678 absorption line. Visually, the He~I $\lambda$6678 absorption in SN 2020acat is weaker and bluer relative to its H$\alpha$ emission than seen in SN 1993J. Through this comparison it is possible to attribute the small excess in the blueward corner of the H$\alpha$ emission of SN 2024ehs to a weak signature from He~I $\lambda$6678 absorption. This suggests that the ``broken'' H$\alpha$ emission typical of Type IIb can take on a wide range of appearances, and this initial comparison may suggest that there is a relation to the relative velocity of the lines.

\section{Methods and Analysis}\label{sec:methods}

\subsection{Velocity evolution}\label{ssec:vels}

\begin{figure}
    \centering
    \includegraphics[width=1\linewidth]{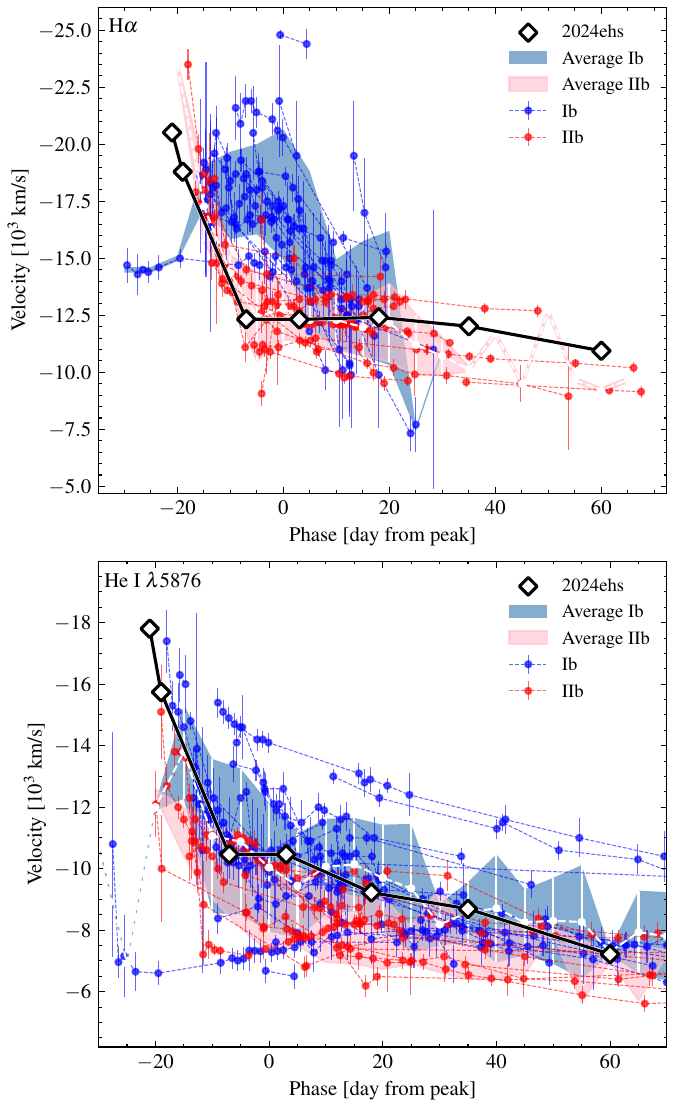}
    \caption{Velocity evolution of the peak absorption of the H$\alpha$ and He $\lambda$5876 lines in the SN 2024ehs spectra, as compared to Ib and IIb data from \cite{Liu2016}. Individual connected dots represent following data points of the same object. The shaded blue and red areas represent the weighted average of Type Ib and IIb objects, respectively.}
    \label{fig:vels_ev}
\end{figure}

\cite{Liu2016} (hereafter L16), performed a sample study on observed differences between Type Ib and IIb spectra, including  the velocity evolution of certain absorption lines. The analysis is constructed from a sample of 14 Type IIb and 21 Type Ib objects, with 504 spectra in total across both types. Fig. \ref{fig:vels_ev} shows the velocity evolution of SN 2024ehs in comparison to the data from L16. Velocities are estimated fitting a parabola to the absorption feature, to match the velocity estimates made by L16. The displayed average of each type is a recreation of the averaging performed in L16, where a re-binning of each individual supernova is performed first, followed by an averaging between the different objects within each bin. Comparison is only possible for the H$\alpha$ and He~I $\lambda$5876 lines, since these are the only lines that can be measured continuously through the spectral evolution of SN 2024ehs. While the sample from L16 shows a slightly lower velocity in Type IIbs for the He~I $\lambda$5876 line, the overlap between the two types is significant, and SN 2024ehs falls within both Types. The H$\alpha$ evolution shows a more clear distinction. Under the assumption that the absorption feature around 6300~$\AA$ in the SN 2024ehs spectra is the result of H$\alpha$, or at least the same feature as being measured by L16, this comparison shows a strong agreement with the average velocity evolution of this line in Type IIb SNe.

\subsection{Helium in SN 2024ehs}\label{ssec:helium}

\subsubsection{Velocity evolution of P-Cygni lines} \label{sssec:p-cyg_ev}

\begin{figure}[!ht]
    \centering
    \includegraphics[width=\linewidth]{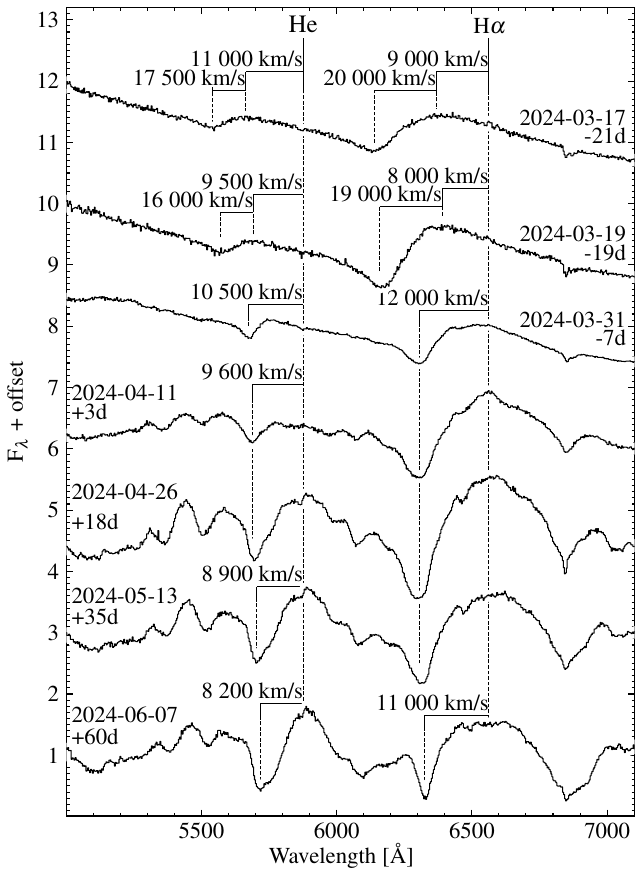} 
    \caption{Closer look at the velocity evolution of H$\alpha$ and He $\lambda$5876 lines. The marked velocities are rounded anchor points for reference and not calculated from any spectral features.}
    \label{fig:vel_ev}
\end{figure}

\begin{figure}[!ht]
    \centering
    \includegraphics[width=\linewidth]{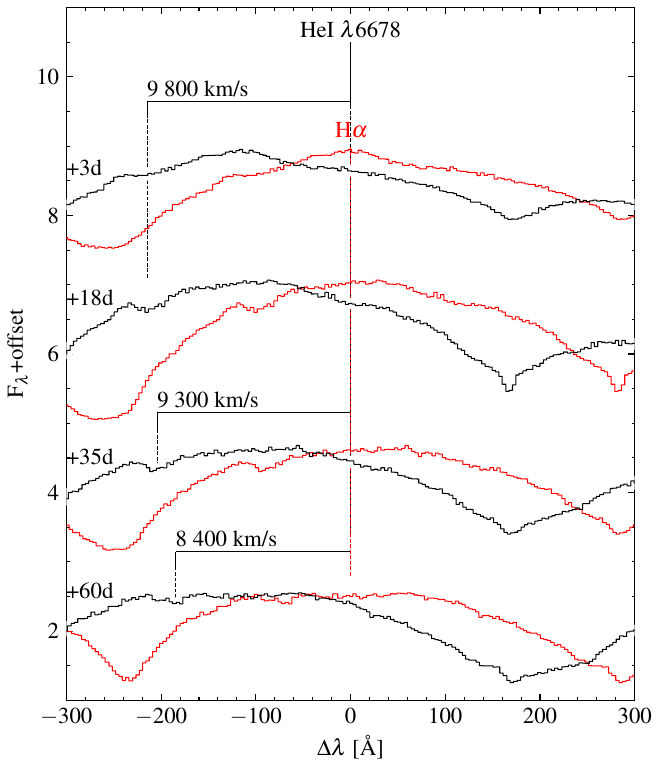}
    \caption{The velocity evolution, assuming that the small absorption emerging to the very left of the H$\alpha$ emission profile is a He~I $\lambda$6678 absorption feature. The wavelength has been shifted to a zero-point of the rest wavelength of He~I $\lambda$6678, and the same spectra, but shifted to the zero point of H$\alpha$ is shown superimposed in red.}
    \label{fig:vel_ev_2}
\end{figure}


Figure \ref{fig:vel_ev} shows a zoomed in spectral evolution focused on the 5000--7000~$\AA$ range, encapsulating the He~I $\lambda$5876 and H$\alpha$ line profiles. Initial P-Cygni lines develop into more complicated structures as more lines show up in the spectra, and the line features are slowing down along the spectral sequence. Fig. \ref{fig:vel_ev_2} shows the velocity evolution of the small feature developing in the H$\alpha$ emission around peak. Superimposed in red is the same spectra centred on the H$\alpha$ line. Here it is assumed that the small feature can be attributed to a He~I $\lambda$6678 absorption, as expected in the spectra of a Type IIb. The velocity evolution of this line may help shed some light into its nature, and what could be causing it to be so weak.  The velocity of this absorption feature is in general agreement with that of He~I $\lambda$5876 in Fig. \ref{fig:vel_ev}, implying these two lines are originating in the same layers of the envelope. This is what we expect from two helium lines, though it is not a definitive confirmation that this is a He~I $\lambda$6678 absorption. Assuming it is, however, also agrees with the weakness of most of the other common helium lines in the spectra of SN 2024ehs. 

\subsubsection{Helium comparison}\label{sssec:helium_comp}

\begin{figure*}[!ht]
   \centering
   \includegraphics[width=\linewidth,scale = 0.5]{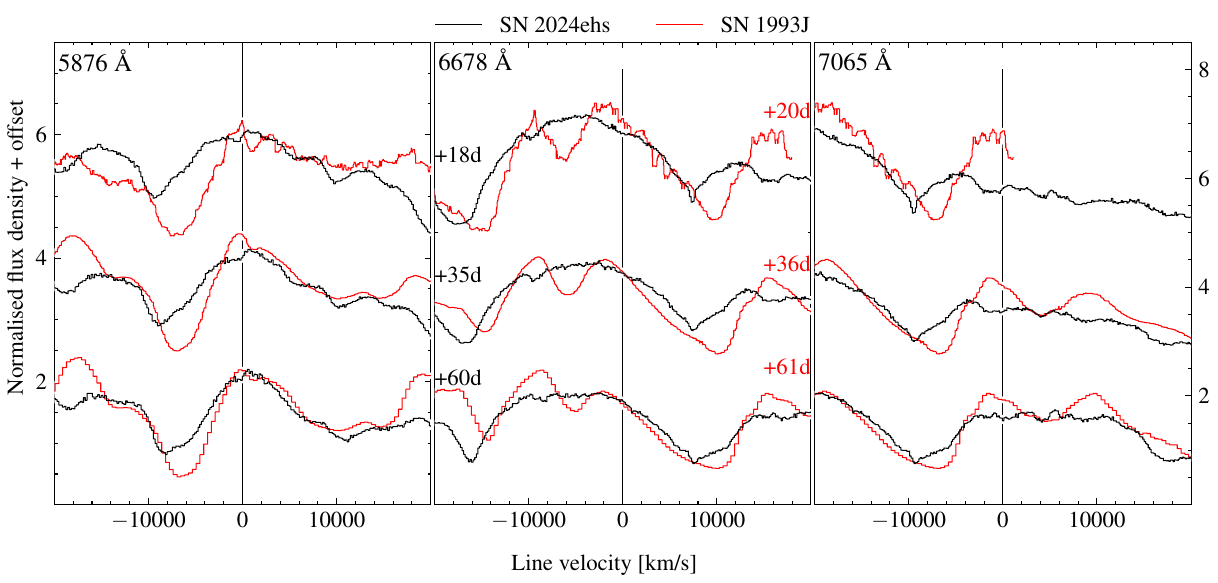}
   \caption{Line evolution of helium lines in the optical. Black lines are spectra from SN 2024ehs at 18, 35 and 60 days post $r$-band peak, red lines are spectra from SN 1993J at 20, 36 and 61 days post $r$-band peak. Each spectrum has been transformed into velocity of individual helium lines, with the respective rest wavelength in the lower right corner of each panel. }
   \label{fig:he_lines}
\end{figure*}

Fig.~\ref{fig:he_lines} shows the velocity evolution of helium lines in the optical spectra. In black are the SN 2024ehs spectra from 18, 35 and 60 days post $r$-band peak, and in red are spectra from SN 1993J, taken at 20, 36 and 61 days post $R$-band peak. Each subplot is centred on an individual helium line, with the wavelength shown in the right-hand corner, and the spectra have been transformed into the velocity of the central line. SN 2024ehs has a faster moving helium envelope that also produces weaker absorptions than those in the corresponding SN 1993J spectra. While this is most evident in the He~I $\lambda$6678 line, the trend can be seen in all helium lines typically seen in Type IIb SNe. The spectra of SN 1993J also show a double-peaked feature around the $\lambda$7065 line, where SN 2024ehs instead displays a shallow, box-like profile. 

\subsection{Spectral Modelling with \texttt{SUMO} }\label{ssec:sumo}

To obtain a better understanding of the ejecta structure and composition of SN 2024ehs, it is insightful to try and match the spectra using spectral modelling efforts. To this end, we make use of the SUpernova MOnte carlo code \citep[ \texttt{SUMO}; ][]{Jerkstrand2011, Jerkstrand2012, Jerkstrand2014}. \texttt{SUMO} is a Monte Carlo radiative transfer code, operating in NLTE. Given an ejecta composition and density structure, \texttt{SUMO} solves for the temperature-, level- and ionisation structures within the nebula, producing a model spectrum. In this subsection, we create model spectra both for our nebular- and (late) photospheric spectra.  

\subsubsection{Nebular modelling}\label{sssec:SUMO_neb}

\begin{figure*}[!ht]
    \centering
    \includegraphics[scale = 0.9]{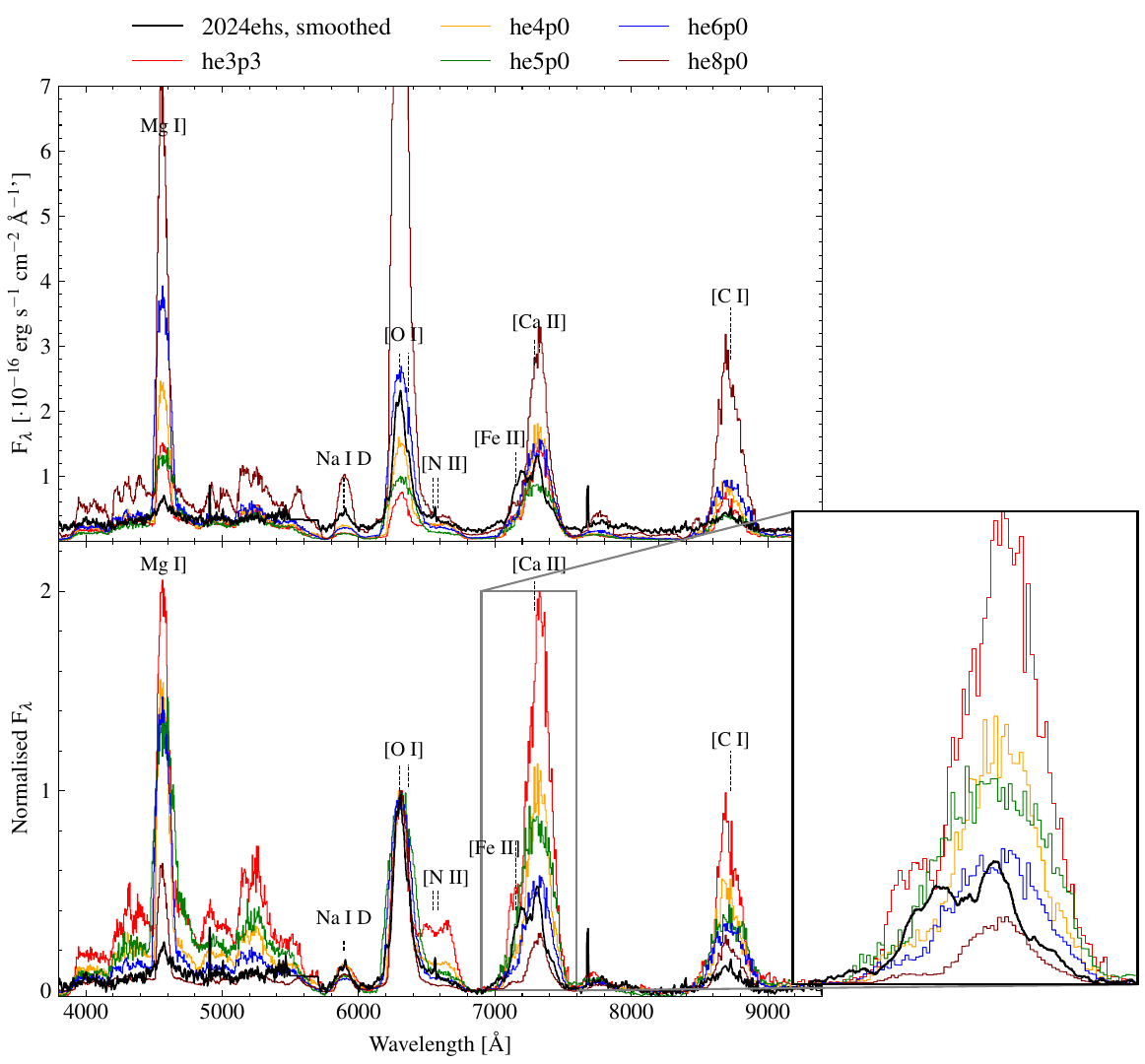}
    \caption{The (smoothed) nebular spectrum of SN 2024ehs, taken $\sim 239$ days post explosion (black) compared to \texttt{SUMO} models from \citet{barmentloo2024}.  The top panel shows the models scaled to a nickel mass of M(Ni) = 0.09 $\mathrm{M}_{\odot}$, and the bottom panel shows both models and spectra scaled between the [O I] $\lambda\lambda$6300, 6364~$\AA$ feature and a baseline. For each spectrum in the bottom panel, the peak of [O I] $\lambda\lambda$6300, 6364~$\AA$ is set to 1 and the mean of the quasi-continuum, as defined by the range between 6800--6900~$\AA$, is set to 0. The most prominent emission lines are marked, and the important [Ca II] feature is singled out.}
    \label{fig:sumo}
\end{figure*}

Fig.~\ref{fig:sumo} shows the nebular spectrum of SN 2024ehs compared to \texttt{SUMO} nebular models from \citet{barmentloo2024}. These models have different pre-explosion masses, nickel masses and mixing values, and are denoted by their helium core mass. The models have been adjusted to allow for comparison with SN 2024ehs according to

\begin{equation}
    F_{m, new} = F_{m,raw}\left(\frac{d_m}{d_{ehs}}\right)^2\left(\frac{M(^{56}Ni)_{ehs}}{M(^{56}Ni)_m}\right)e^{2(t_m-t_{ehs})/{\tau_{Co}}},
    \label{eq:sumo}
\end{equation}
where $d_m = 10$ Mpc and $d_{ehs} = 23.8$ Mpc is the distance set for the models and that to SN 2024ehs, respectively, $\mathrm{M}(^{56}Ni)_{ehs} = 0.09$ $\mathrm{M}_{\odot}$ is an estimation for the nickel mass and $t_{ehs} = 239$ d is the time since explosion of SN 2024ehs, $\mathrm{M}(^{56}Ni)_m$ is the nickel mass for the individual model, and $\tau_{{Co}} = 111$ d is the decay time of $^{56}\mathrm{Co}$. We note that this rescaling is independent on uncertainties on $\mathrm{M}(^{56}Ni)_{ehs}$ and $d_{ehs}$, as the $\mathrm{M}(^{56}Ni)_{ehs}$ estimate is directly proportional to $d_{ehs}^{2}$ . The e-folding time ($\tau_{Co}/2$) is chosen to match the slope of the light-curve radioactive tail to that of the $^{56}\mathrm{Co}$ decay (see Sect.~\ref{ssec:photometry}). The nickel mass of SN 2024ehs has not been explicitly investigated, and the arbitrary estimation 0.09 $\mathrm{M}_{\odot}$ is based on the light curve placement. The absolute magnitude is driven by the nickel mass, and in Fig. \ref{fig:comp_lc}, SN 2024ehs is between SN 1993J \citep[with a nickel mass of $\sim0.073~\mathrm{M}_{\odot}$, ][]{Blinnikov1998} and 2020acat \citep[$\sim0.13~\mathrm{M}_{\odot}$, ][]{ergon2023}, and thus would likely have a nickel mass somewhere between these values.

The strengths of the oxygen lines are strongly dependent on the progenitor mass, and thus the [O I] $\lambda\lambda$6300,6364 doublet is the most important line for mass estimation of the progenitor \citep{Jerkstrand2015}. However, both distance and nickel mass affect the scaling of the spectrum (see Equation \ref{eq:sumo}). Therefore, we also scale each spectrum to a normalised range, according to 
\begin{equation}
    x_{new} = \frac{x - x_{min}}{x_{max}-x_{min}}, 
\end{equation}
where $x_{max}$ is the peak of the [O I] $\lambda\lambda$6300, 6364 doublet and $x_{min}$ is the average of the quasi-continuum for each spectrum (bottom panel in figure \ref{fig:sumo}).  Most importantly (and evident in the figure), the ratio between the [O I] $\lambda\lambda$6300, 6364 and [Ca II]$\lambda\lambda$7291, 7323 doublets has a mass dependence (\citealp{Fransson1989}; but see also \citealp{Jerkstrand2017} for caveats), so with the [O I] line set we can look at how the [Ca II] line matches our spectrum between the models. 

In both panels, the SN 2024ehs spectrum is in best agreement with the \texttt{He6p0} model, which is the $6~\mathrm{M}_{\odot}$ Helium core, with a zero-age main sequence (ZAMS) mass of $23.3~\mathrm{M}_{\odot}$, and a pre-explosion core mass of $4.45~\mathrm{M}_{\odot}$. In the lower panel, the [O I]/[Ca II] ratio is consistent with the \texttt{He6p0} model, strengthening the argument. Furthermore, the strength of the [N II] $\lambda$ 6548, 6583 doublet \citep[another progenitor mass diagnostic,][]{barmentloo2024} is also well matched to the \texttt{He6p0} model, something that may not be said for the lower-mass \texttt{He3p3} and \texttt{He4p0} models. The agreement with \texttt{He6p0} and its correct scaling in the upper panel implies that a nickel mass of 0.09 $\mathrm{M}_{\odot}$ is a reasonable estimate for this object, though that assumes a correct distance estimate. In this model comparison we can again see that we have an unusually strong iron line at $\lambda$7155, as evident in the comparison to other objects in Sec.~\ref{ssec:spectra}.

\subsubsection{Photospheric modelling}\label{SUMO_phot}

\begin{figure*}[ht]
    \centering
    \includegraphics[width=\linewidth]{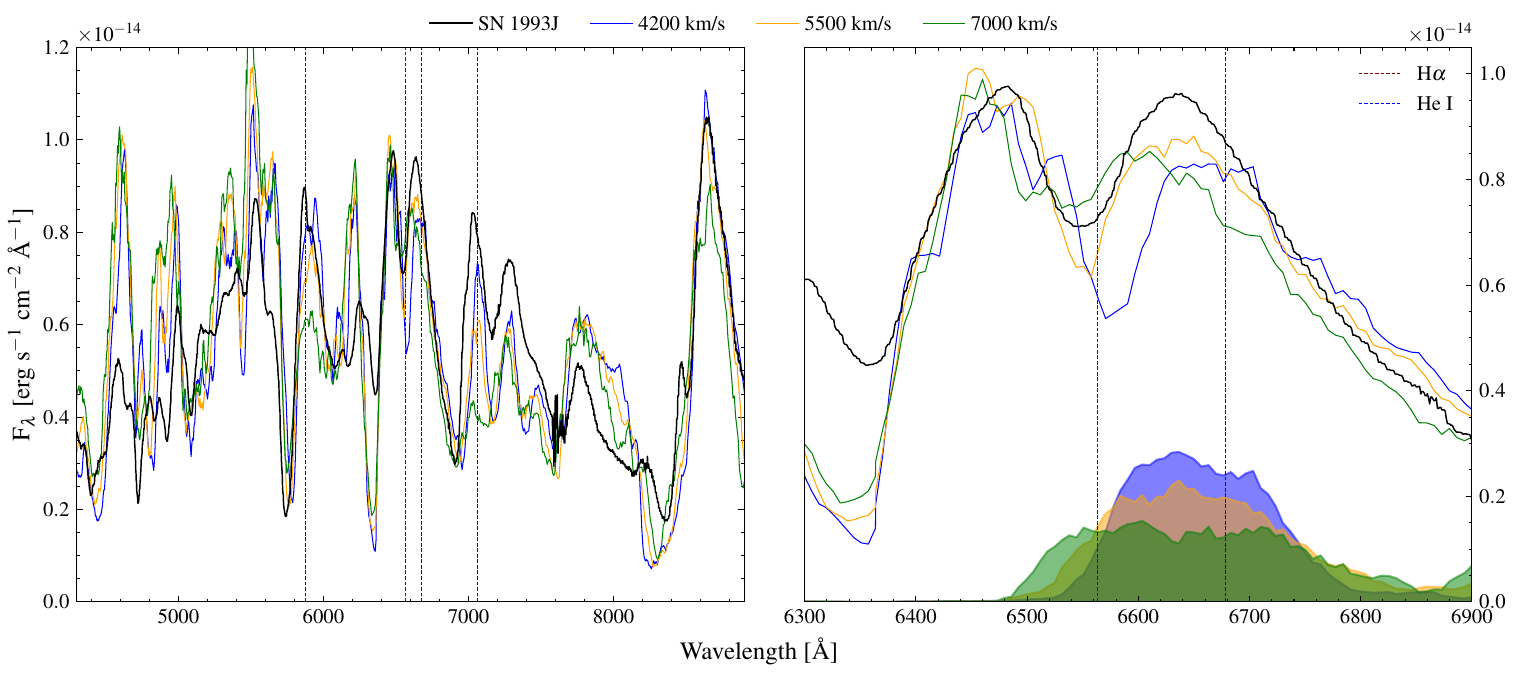}
    \caption{Comparison between the spectrum of SN 1993J 55 days after explosion with SUMO models at 40 days after explosion, at three different core velocities. The right hand panel shows the same comparison, zoomed into the H$\alpha$/He I 6678 range, with the contribution from helium in each spectrum filled in at the bottom. }
    \caption*{\footnotesize Note: To best match the $H\alpha$ + He I 6678 complex in the SN 1993J 55d spectrum, it was found that the best fitting nebular models had to be evaluated at 40 days post explosion, indicating that the model had lower helium envelope densities than SN 1993J}
    \label{fig:93j_models}
\end{figure*}

Now we turn our attention to our late photospheric spectra, and in particular to the weak feature in the H$\alpha$ emission discussed in Section \ref{sssec:p-cyg_ev}. To study what influences the strength of this (most likely He-) feature, we created a small grid of four \textsc{SUMO} models. In these models, we vary the velocity of the lowest velocity He-rich material, \vhe, between 4 200, 5 500, 7 000 and 9 000 \kms. The photospheric modelling is explained in greater detail in Appendix \ref{app:photosp}. In short, the 4 200 \kms\ model was first tuned to SN 1993J (one of the most well-documented Type IIb SNe), and using this model as a base, we created three more variants where only \vhe is varied.

\begin{figure*}[!ht]
    \centering
    \includegraphics[scale = 0.85]{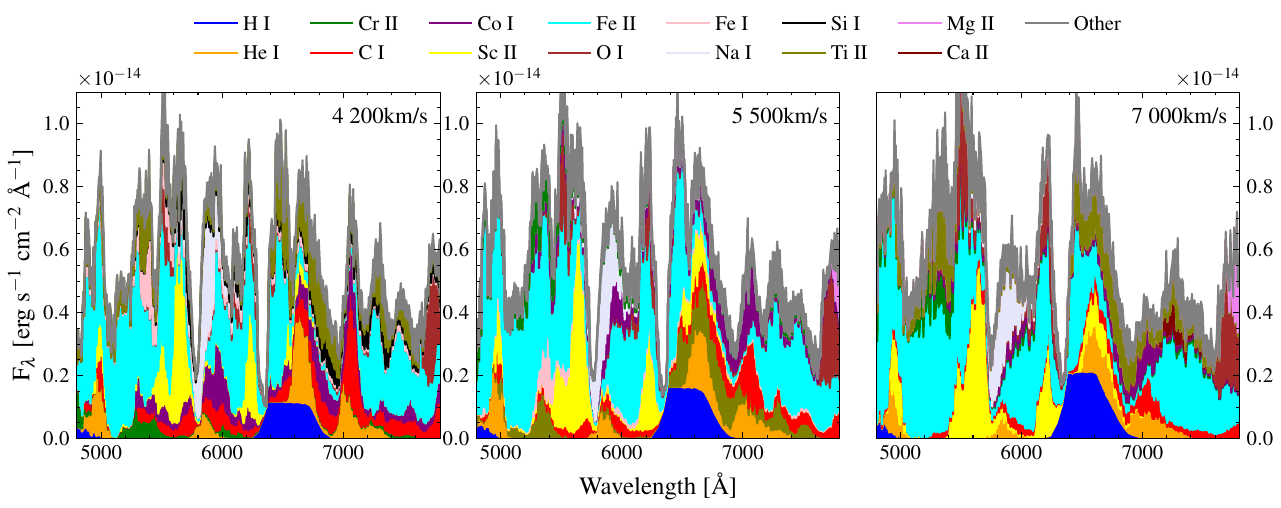}
    \caption{The total flux from the \textsc{SUMO} models with core velocities of 4 200, 5 500 and 7 000 \kms, with the contribution from the 12 most prominent species filled in.}
    \label{fig:sumo_phot}
\end{figure*}

\begin{figure*}
    \centering
    \includegraphics[width=\linewidth]{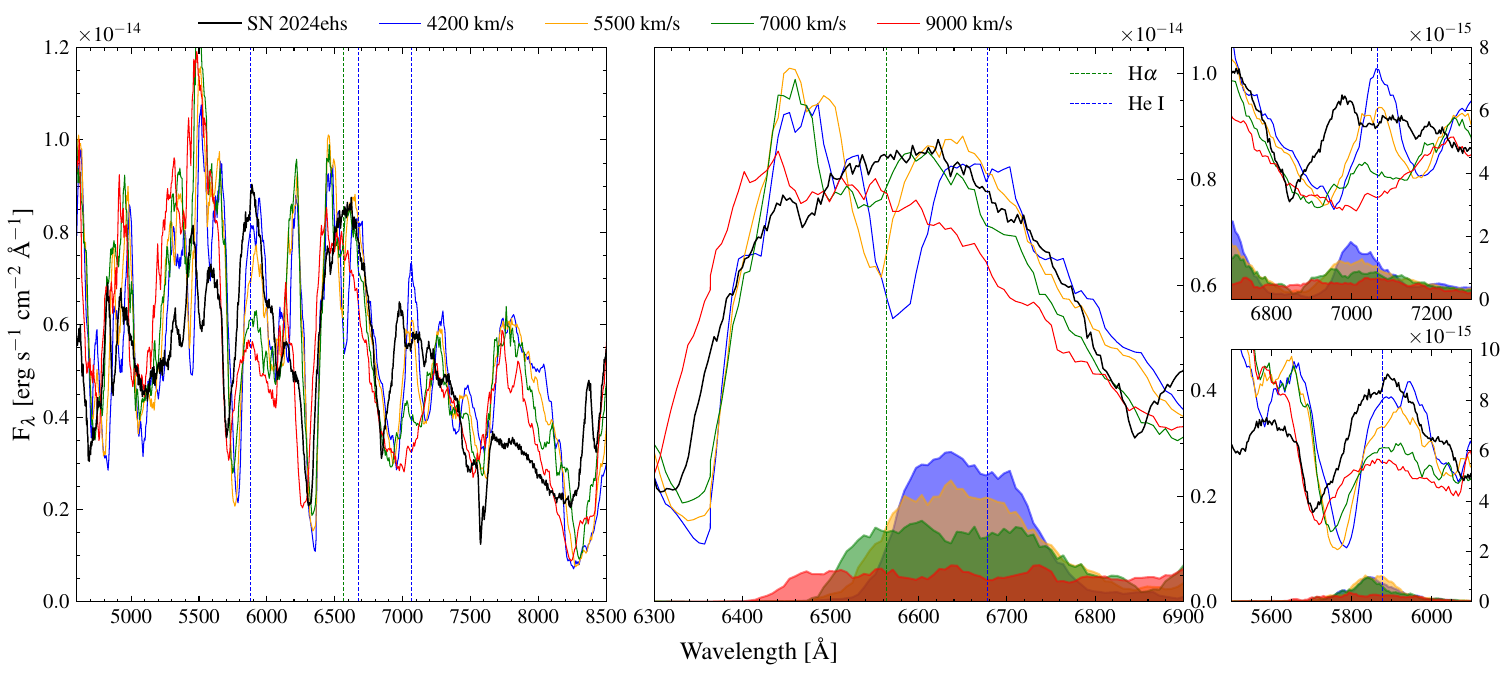}
\caption{Same as Fig \ref{fig:93j_models} but comparing the SUMO models (at 40 days post explosion) to SN 2024ehs at $\sim56$ days post explosion. Here a model with core velocity at 9 000 \kms\ is added, to further demonstrate the effect the increase in velocity has on the spectra. The smaller panels focus in on the area around He I $\lambda$6678 and He I $\lambda$ 7065 lines. The blue dotted line signifies the rest wavelength of the helium line in focus, while the green dotted line in the bigger panel is the rest wavelength of H$\alpha$.}
    \label{fig:24ehs_models}
\end{figure*}


First comparing the model spectra to SN 1993J on the whole (Fig.~\ref{fig:93j_models}), we can say that the models reproduce SN 1993J satisfactorily in broad strokes, with the majority of the spectra matching within $25\%$ to SN 1993J. The most obvious discrepancies occur around blueward emission lines, at wavelengths of $\sim$ 3500, 3800, 4000 and 4600 Å. The ion-by-ion decompositions (Fig.~\ref{fig:sumo_phot}) show that these are mostly caused by overestimations of heavier atoms, in particular Ca II, Cr II, Ti II and Sc II, suggesting that the innermost zones of the \texttt{He6p0} model are too massive to match SN 1993J. 

With this satisfactory match, we can further investigate the \halpha + He I $\lambda$6678 complex, of which a zoom-in is shown in the right-hand panel of Figure \ref{fig:93j_models}. The emission part of \halpha is well reproduced by the models, whereas the absorption is too strong, especially at higher velocities. More interestingly, the He I $\lambda$6678 profile clearly shifts in strength and in the location of its minimum with increasing \vhe. The reader is reminded that this is only due to density and velocity effects; the He mass is exactly the same ($\sim$ 1.0 \msun) in all of these models. Firstly, the increased expansion velocities of the helium-rich material lead to lower densities and thus lower optical depths for the helium lines, causing the depth of the absorption profile to decrease. Secondly, larger \vhe spreads out the emission component over a broader wavelength range (see the helium-only contribution in Figure \ref{fig:93j_models}). 

Focusing more explicitly now on the weak feature in this region in SN 2024ehs, the right-hand panel of Figure \ref{fig:24ehs_models} shows a comparison of the 56-day SN 2024ehs spectrum to the \texttt{v4200} -- \texttt{v9000} models (again, for the model descriptions see Appendix \ref{app:photosp}). Again, as clearly illustrated by the helium-only contributions for these two models, the increase in \vhe to 9 000 \kms spreads out the emission component so much, that it is barely noticeable in the total model spectrum, despite this model having equal helium mass as the \texttt{v4200} model. Interestingly, the weak absorption feature around 6450 $\AA$ in the SN 2024ehs spectrum is well-matched by the \texttt{v9000} model, further strengthening its identification as weak He I $\lambda$6678 absorption. Considering the other He lines, an increase in velocity also leads to a better match for He I $\lambda$5876, where in the models the majority of the feature is in fact due to Na I rather than He I. For He I $\lambda$7065, the match worsens, but interpreting the absorption minimum is made difficult by the neighbouring telluric.

The findings presented here agree with earlier results in the literature in the discussion on how much helium can effectively be hidden in a Type Ic SN, of which the progenitors are typically thought of as stripped of their helium layer. In these works, it is found that Type Ic SN spectra can be reproduced with presence of a non-negligible amount of helium; \citet{Hachinger2012} and \citet{Williamson2021} find that when adding successively more massive helium shells onto a C+O -core star, this 'hideable' helium mass is around 0.05 -- 0.1 \msun. \citet{Dessart2012} emphasise that for the non-thermal electrons (crucial for production of He I lines) to reach the helium atoms, mixing of $^{56}$Ni into the helium-rich layers is essential, and that the lack of helium lines is thus not necessarily an indication of a lack of helium in the ejecta. Our models show that density effects can be another cause of a lack of strong optical helium lines (hiding $\sim$1 \msun of helium a few weeks post maximum light) which is of particular interest for Type Ic-BL SNe with their characteristically large expansion velocities. 




\section{Discussion}\label{sec:discussion}

\subsection{SN 2024ehs as a Type IIb supernova}

There are two published classifications for SN 2024ehs, one made two days after discovery, on 2024-03-17 \citep{Li2024} and one on 2024-03-31 \citep{Leadbeater2024}, 16 days after discovery. Both report a classification of Type II, while mentioning the possibility of a development into a Type IIb. \cite{Leadbeater2024} mentions the still prominent hydrogen lines and lack of strong helium lines to reject a Type IIb classification at the time. However with the further development of the spectra, it is now clear that the strength of the hydrogen lines do not continue increasing post peak. While it is unfortunate that a part of the transition between photospheric and nebular phase is unavailable, the nebular spectrum taken 216 days post peak no longer contains any signs of hydrogen. 

Although a few individual features make the spectra of SN 2024ehs stand out, its overall spectral shape and evolution closely match those of the Type IIb comparison objects: from an initial spectrum with weak but present helium lines, through a photospheric phase with a broadening H$\alpha$ and deepening helium lines, to a nebular spectrum dominated by [O I] and [Ca II]. No two objects are identical in all regards, but SN 2024ehs displays enough key features for us to classify it as a Type IIb.  

\subsection{Properties of Stripped-Envelope Supernovae}

In comparison with SESNe (Fig. \ref{fig:comp_lc}), the light curve of SN 2024ehs is narrow, with a fast decline rate and no visible shock breakout. This may indicate a low ejected mass \citep{Arnett1982}, relatively compact progenitor, no noticeable surrounding shells of material \citep{falk1977} and relatively low $\gamma$-trapping \citep{jerkstrand2025corecollapsesupernovae}. The compactness of the progenitor may also indicate it as a candidate for a WR progenitor \citep{Chevalier2010}.

In velocity comparisons with \cite{Liu2016}, SN 2024ehs has a He~I $\lambda 5876$ absorption velocity significantly higher than the typical Type IIb. Type Ib SNe typically reach higher velocities in the helium lines, and the similarity with Type Ib velocities would again support the idea that SN 2024ehs has a low ejecta mass. The velocity evolution of the H$\alpha$ line is in agreement with the average evolution for Type IIb. 

We understand the progenitors of SESNe to be massive stars, likely in a binary system experiencing heavy mass loss \citep{BranchWheeler}. The nebular spectrum of SN 2024ehs matches a \textsc{SUMO} model evolved with an original mass of $23.3~M_{\odot}$, though this is not a unique correlation with the final pre-explosion mass \citep{Dessart2023}. The final pre-explosion core mass has a more direct impact on the spectra, and here we do see that SN 2024ehs has a pre-explosion core mass of 4.45 \msun. 

\subsection{Insights from radiative transfer modelling}

The nebular spectral comparison to \textsc{SUMO} models finds a best match for SN 2024ehs to the \texttt{He6p0} model, when comparing both the ratio between the oxygen and calcium doublets as well as the strength of the nitrogen doublet. The fit to the \texttt{He6p0} model is used to support a $^{56}$Ni mass around $0.09~\mathrm{M}_{\odot}$, as used in the comparison not scaled to the oxygen line. However, we once again remind the reader of the uncertainty in the distance, complicating the $^{56}$Ni estimate from spectra alone. 

The photospheric comparisons using \textsc{SUMO} provide an interesting insight into how velocities impact the strength and shapes of lines. For SN 2024ehs, we see that higher velocities correlate with weaker lines from helium. The explanation for this is that the optical depth of these lines decreases significantly, lowering the scattering and absorption effects. The lack of strength in the helium lines does thus not necessarily mean a lack of helium, and the correlation of absorption depth with line velocity is strengthened by both model and observational comparisons in this work. 
Starting with a photospheric model that is in general agreement with SN 1993J, we confirm that changing only \vhe distorts the spectral features into something closer to the behaviour we see in SN 2024ehs.
Another interesting aspect of the photospheric \textsc{SUMO} modelling is the broad, boxy profile in the spectrum of SN 2024ehs around the He~I $\lambda$7065 line, that is not reproduced by the model. While high velocities contribute to the blending of lines, the models imply that the peak on the blue side should disappear instead of spreading out. It is possible that other lines are contributing here, e.g. stronger iron than found in the models. Such an excess of iron in the ejecta of SN 2024ehs would agree with the unusually strong iron emission observed in the nebular spectra. 

To complement the \textsc{SUMO} modelling, we can go back to the spectral comparisons (see Fig. \ref{fig:spec_comp}) and look at the objects SN 1993J and 2020acat. In section \ref{ssec:spectra}, it is noted that SN 2020acat is in-between SN 2024ehs and SN 1993J with regards to the shape of the H$\alpha$/He~I $\lambda$6678 profile. Upon further inspection, SN 2020acat also seems to have a sort of middle ground with regards to this He$\lambda$7065 environment; something that initially could be described as a boxy profile, but with a hint of two distinct lines. Looking at the nebular spectrum of SN 2020acat, there is also an extension of the [Ca II] doublet to the blue side, which may be a signal from the same iron line that is so strong in SN 2024ehs. This would again place SN 2020acat between the extremes of SN 2024ehs and SN 1993J, indicating a more diverse spread within SESNe. 

\section{Conclusion}\label{sec:conclusion}
This paper presents data on the Type IIb SN 2024ehs, discovered in March 2024.
Photometric data are presented in the $r$, $g$ and $o$ bands, as well as spectroscopic data in the optical ($\sim 3000-9500$~$\AA$). We have studied the behaviour of certain helium lines and have made comparisons to similar objects and to numerical models in order to attain an understanding of their emergence. 

Comparisons of our nebular spectrum to NLTE \textsc{SUMO} spectral models suggests that the progenitor of SN 2024ehs had a relatively high helium core mass ($\sim4.5~M_{\odot}$ at the time of explosion), but this cannot reliably be extrapolated to a ZAMS mass. Perhaps most interestingly, \textsc{SUMO} modelling of our late photospheric spectra shows that higher ejecta velocities dampen the emission of certain lines due to density/opacity effects. In the spectral evolution of SN 2024ehs, we see high velocities in weak helium lines, and our simulations suggest that this weakness may not be due to a lack of helium, but rather to lower densities in He-rich layers. This finding is in line with earlier studies which show that non-negligible amounts of helium may be hidden in Type Ic SNe. With the comparison to another Type IIb, SN 2020acat, it becomes clear that these helium lines showcase a range of different strengths and velocities, solidifying the correlation with higher velocities displaying weaker lines. 

SN 2024ehs, despite being a Type IIb SN, thus provides another example of the difficulty in distinguishing between Type Ib and Type Ic SNe when relying on optical spectra alone. The advent of near-infrared (NIR) facilities such as the James Webb Space Telescope (JWST) provides the opportunity to instead look for the He I $\lambda$ 1.08 and He I $\lambda$ 2.06 $\mu$m lines, which do provide a definitive answer to the question of presence of helium in the ejecta \citep{vanBaal2024}.

\begin{acknowledgements}

This work was conducted as part of a Master’s Thesis project at Stockholm University. W. Sand Hellman gratefully acknowledges the support and contributions of the members of the Department of Astronomy, whose involvement was instrumental in the completion of this research. S. J. Brennan acknowledges funding by the European Union (ERC, project number 101042299, TransPIre). Views and opinions expressed are however those of the author(s) only and do not necessarily reflect those of the European Union or the European Research Council Executive Agency. Neither the European Union nor the granting authority can be held responsible for them. SB and AJ acknowledge funding from the Swedish National Research Council (Starting Grant 2018–03799). The computations in this work
were enabled by resources provided by the Swedish National
Infrastructure for Computing (SNIC), the National Academic Infrastructure for Supercomputing in Sweden (NAISS), and at the Parallelldatorcentrum (PDC) Center for High Performance Computing, Royal Institute of Technology (KTH), partially funded by the Swedish Research Council through grant agreements nos 2022–06725 and 2018–05973.

This research has made use of NED, which is operated by the Jet Propulsion Laboratory, California Institute of Technology,
under contract with the National Aeronautics and Space Administration.  Based on observations obtained with the Samuel Oschin Telescope 48-inch and the 60-inch Telescope at the Palomar Observatory as part of the Zwicky Transient Facility project. ZTF is supported by the National Science Foundation under Award \#2407588 and a partnership including Caltech, USA; Caltech/IPAC, USA; University of Maryland, USA; University of California, Berkeley, USA; University of Wisconsin at Milwaukee, USA; Cornell University, USA; Drexel University, USA; University of North Carolina at Chapel Hill, USA; Institute of Science and Technology, Austria; National Central University, Taiwan, and OKC, University of Stockholm, Sweden. Operations are conducted by Caltech's Optical Observatory (COO), Caltech/IPAC, and the University of Washington at Seattle, USA. 
\end{acknowledgements}

\appendix

\section{Photospheric modelling}\label{app:photosp}
Here we describe how the late photospheric \textsc{SUMO} models were created. 

As a base, we used the same helium star model grid as described in \citet{barmentloo2024} (which originate from \citet{Woosley2019} and \citet{Ertl2020}). In particular, the \texttt{He6p0} model was adopted. To this helium star, we attach a H-rich envelope of mass 0.15 \msun, with elemental composition as in \citet{Woosley_1994} (believed to be typical for the hydrogen envelopes of Type IIb progenitors). To obtain as close a match as possible to the spectra of SN 1993J, we set the inner velocity of the H-rich envelope to 9 000 \kms, while the outer velocity was set to 20 000 \kms. The outer velocity of the macroscopically-mixed core (and thus the inner velocity of the He-rich envelope) was then varied between 4 200 (our best estimate for SN 1993J), 5 500 and 7 000 \kms to investigate the formation of the He I $\lambda$6678 feature (see Section \ref{sssec:p-cyg_ev}). To enforce these two velocity constraints, it was necessary to change the density structures of the original \citet{Ertl2020} He/C and He/N envelopes. In line with previous work \citep{Jerkstrand2015}, density profiles were defined as $\rho = \rho_{0} \times v^{-n}$ power laws, with the power indices $n$ at -4, -6 and -8 for the He/C-, He/N- and H-envelopes, respectively. For each profile, $\rho_{0}$ (as well as the velocity of the He/C - He/N zone boundary) is then set by the mass in the respective zones (which was kept the same as in the original \citet{Ertl2020} models). An exploration of the effect of the choice of power indices was performed, and it was found that these did not appreciably influence the resulting spectra. 

Besides these three models (which we dub \texttt{v4200}, \texttt{v5500}, \texttt{v7000}), we created a fourth model, \texttt{v9000}, where the inner velocity of the H-rich envelope was set to 12 000 \kms, and the inner velocity of the He-rich envelope was set to 9 000 \kms. These values were chosen specifically to try and reproduce the \halpha + He I $\lambda$6678 complex as seen in SN 2024ehs. 

\section{SEDM spectra}

\begin{figure*}[h!]
    \centering
    \includegraphics[width=\linewidth]{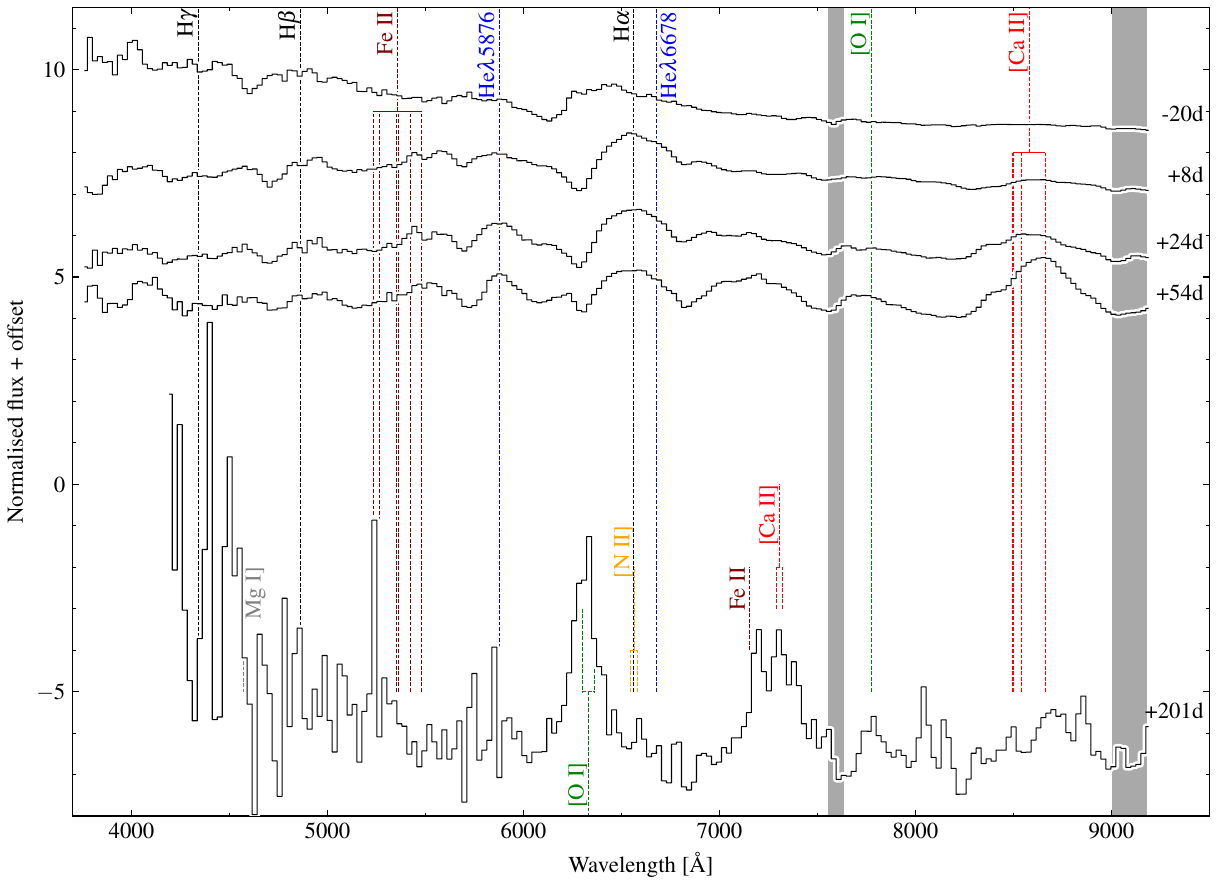}
    \caption{Spectra of SN 2024ehs taken with the SEDM on the Palomar 60-inch (P60) telescope. The marked days are in reference to r-band peak. The nebular spectrum has been slightly cut around 4000\AA to reduce overlap.}
    \label{fig:sedm}
\end{figure*}

\section{Tables}\label{app:tables}

\onecolumn

\begin{longtable}{lccc}
\caption{Lightcurve photometry for SN\,2024ehs in $g$, $r$, and $o$ bands (first 10 and last 10 measurements shown). The full table is available online. Limiting magnitudes are indicated with $<$ symbol.} \\
\toprule
MJD & Filter & Magnitude & Error \\
\midrule
\endfirsthead

\caption{\tablename\ \thetable{} -- continued from previous page} \\
\toprule
MJD & Filter & Magnitude & Error \\
\midrule
\endhead

\midrule
\multicolumn{4}{r}{Continued on next page} \\
\midrule
\endfoot

\bottomrule
\endlastfoot

60384.50 & o & 18.14 & $\pm$ 0.08 \\
60384.50 & o & 18.17 & $\pm$ 0.08 \\
60384.50 & o & 18.08 & $\pm$ 0.06 \\
60386.40 & o & 16.98 & $\pm$ 0.03 \\
60386.40 & o & 16.95 & $\pm$ 0.03 \\
60386.40 & o & 17.00 & $\pm$ 0.03 \\
60386.40 & o & 16.93 & $\pm$ 0.03 \\
60387.40 & o & 16.85 & $\pm$ 0.03 \\
60387.40 & o & 16.75 & $\pm$ 0.02 \\
60387.40 & o & 16.85 & $\pm$ 0.03 \\
-- & -- & -- & -- \\
-- & -- & -- & -- \\
-- & -- & -- & -- \\
60628.50 & r & 19.38 & $\pm$ 0.18 \\
60628.50 & g & 20.28 & $\pm$ 0.19 \\
60635.50 & r & < 19.56 & -- \\
60635.50 & g & < 20.17 & -- \\
60646.50 & g & 20.74 & $\pm$ 0.17 \\
60646.50 & g & < 20.35 & -- \\
60646.50 & r & 19.87 & $\pm$ 0.16 \\
60646.50 & r & 19.76 & $\pm$ 0.07 \\
60646.50 & r & 19.87 & $\pm$ 0.16 \\
60648.50 & r & 19.85 & $\pm$ 0.14 \\

\end{longtable}
\label{tab:sn2024ehs_lightcurve_trunc}

\twocolumn

\onecolumn

\begin{longtable}{l l l l l}
\caption{Complete spectroscopic observations of SN\,2024ehs. Wavelength ranges are given after MW extinction and redshift corrections.} \\
\toprule
MJD & Date & Instrument & Telescope & Wavelength range \\
\midrule
\endfirsthead

\caption{\tablename\ \thetable{} -- continued from previous page} \\
\toprule
MJD & Date & Instrument & Telescope & Wavelength range \\
\midrule
\endhead

\midrule
\multicolumn{5}{r}{Continued on next page} \\
\endfoot

\bottomrule
\endlastfoot

60386.94 & 2024-03-17 & Nordic Optical Telescope  & ALFOSC & 3837--9738\,\AA \\
60388.94 & 2024-03-19 & Nordic Optical Telescope  & ALFOSC & 3837--9748\,\AA \\
60400.97 & 2024-03-31 & Nordic Optical Telescope  & ALFOSC & 3489--9726\,\AA \\
60411.98 & 2024-04-11 & Nordic Optical Telescope  & ALFOSC & 3489--9595\,\AA \\
60426.99 & 2024-04-26 & Nordic Optical Telescope  & ALFOSC & 3488--9748\,\AA \\
60443.90 & 2024-05-13 & Nordic Optical Telescope  & ALFOSC & 3490--9607\,\AA \\
60468.88 & 2024-06-07 & Nordic Optical Telescope  & ALFOSC & 3487--9638\,\AA \\
60620.22 & 2024-11-06 & Nordic Optical Telescope  & ALFOSC & 3487--9610\,\AA \\
60624.53 & 2024-11-10 & Palomar 5.1m Hale  & DBSP & 3388--10459\,\AA \\

\end{longtable}
\label{tab:sn2024ehs_spectra_full}

\twocolumn

\pagebreak

\bibliography{references}{}

@ARTICLE{Mandigo-Stoba2022,
       author = {{Mandigo-Stoba}, Milan Sharma and {Fremling}, Christoffer and {Kasliwal}, Mansi},
        title = "{DBSP\_DRP: A Python package for automated spectroscopic data reduction of DBSP data}",
      journal = {The Journal of Open Source Software},
         year = 2022,
        month = feb,
       volume = {7},
       number = {70},
          eid = {3612},
        pages = {3612},
          doi = {10.21105/joss.03612},
archivePrefix = {arXiv},
       eprint = {2107.12339},
 primaryClass = {astro-ph.IM},
       adsurl = {https://ui.adsabs.harvard.edu/abs/2022JOSS....7.3612M}
}

@article{Coughlin2023,
	title        = {{A Data Science Platform to Enable Time-domain Astronomy}},
	author       = {{Coughlin}, Michael W. and {Bloom}, Joshua S. and {Nir}, Guy and {Antier}, Sarah and {du Laz}, Theophile Jegou and {van der Walt}, St{\'e}fan and {Crellin-Quick}, Arien and {Culino}, Thomas and {Duev}, Dmitry A. and {Goldstein}, Daniel A. and {Healy}, Brian F. and {Karambelkar}, Viraj and {Lilleboe}, Jada and {Shin}, Kyung Min and {Singer}, Leo P. and {Ahumada}, Tom{\'a}s and {Anand}, Shreya and {Bellm}, Eric C. and {Dekany}, Richard and {Graham}, Matthew J. and {Kasliwal}, Mansi M. and {Kostadinova}, Ivona and {Kiendrebeogo}, R. Weizmann and {Kulkarni}, Shrinivas R. and {Jenkins}, Sydney and {LeBaron}, Natalie and {Mahabal}, Ashish A. and {Neill}, James D. and {Parazin}, B. and {Peloton}, Julien and {Perley}, Daniel A. and {Riddle}, Reed and {Rusholme}, Ben and {van Santen}, Jakob and {Sollerman}, Jesper and {Stein}, Robert and {Turpin}, D. and {Wold}, Avery and {Amat}, Carla and {Bonnefon}, Adrien and {Bonnefoy}, Adrien and {Flament}, Manon and {Kerkow}, Frank and {Kishore}, Sulekha and {Jani}, Shloke and {Mahanty}, Stephen K. and {Liu}, C{\'e}line and {Llinares}, Laura and {Makarison}, Jolyane and {Olli{\'e}ric}, Alix and {Perez}, In{\`e}s and {Pont}, Lydie and {Sharma}, Vyom},
	year         = 2023,
	month        = {aug},
	journal      = {\apjs},
	volume       = 267,
	number       = 2,
	pages        = 31,
	doi          = {10.3847/1538-4365/acdee1},
	eid          = 31,
	archiveprefix = {arXiv},
	eprint       = {2305.00108},
	primaryclass = {astro-ph.IM},
	adsurl       = {https://ui.adsabs.harvard.edu/abs/2023ApJS..267...31C}
}

@article{pypeit:joss_arXiv,
	title        = {{PypeIt: The Python Spectroscopic Data Reduction Pipeline}},
	author       = {{Prochaska}, J. Xavier and {Hennawi}, Joseph F. and {Westfall}, Kyle B. and {Cooke}, Ryan J. and {Wang}, Feige and {Hsyu}, Tiffany and {Davies}, Frederick B. and {Farina}, Emanuele Paolo},
	year         = 2020,
	month        = {may},
	journal      = {arXiv e-prints},
	pages        = {arXiv:2005.06505},
	eid          = {arXiv:2005.06505},
	archiveprefix = {arXiv},
	eprint       = {2005.06505},
	primaryclass = {astro-ph.IM},
	adsurl       = {https://ui.adsabs.harvard.edu/abs/2020arXiv200506505P}
}

@article{Walt2019,
	title        = {{SkyPortal: An Astronomical Data Platform}},
	author       = {{van der Walt}, St{\'e}fan and {Crellin-Quick}, Arien and {Bloom}, Joshua},
	year         = 2019,
	month        = {may},
	journal      = {The Journal of Open Source Software},
	volume       = 4,
	number       = 37,
	pages        = 1247,
	doi          = {10.21105/joss.01247},
	eid          = 1247,
	adsurl       = {https://ui.adsabs.harvard.edu/abs/2019JOSS....4.1247V}
}

@article{Brennan2022d,
	title        = {{The Automated Photometry of Transients pipeline (AUTOPHOT)}},
	author       = {{Brennan}, S.~J. and {Fraser}, M.},
	year         = 2022,
	month        = {nov},
	journal      = {\aap},
	volume       = 667,
	pages        = {A62},
	doi          = {10.1051/0004-6361/202243067},
	eid          = {A62},
	archiveprefix = {arXiv},
	eprint       = {2201.02635},
	primaryclass = {astro-ph.IM},
	adsurl       = {https://ui.adsabs.harvard.edu/abs/2022A&A...667A..62B}
}

@article{Shingles2021a,
	title        = {{Release of the ATLAS Forced Photometry server for public use}},
	author       = {{Shingles}, L. and {Smith}, K.~W. and {Young}, D.~R. and {Smartt}, S.~J. and {Tonry}, J. and {Denneau}, L. and {Heinze}, A. and {Weiland}, H. and {Flewelling}, H. and {Stalder}, B. and {Clocchiatti}, A. and {F{\"o}rster}, F. and {Pignata}, G. and {Rest}, A. and {Anderson}, J. and {Stubbs}, C. and {Erasmus}, N.},
	year         = 2021,
	month        = {jan},
	journal      = {Transient Name Server AstroNote},
	volume       = 7,
	pages        = {1--7},
	adsurl       = {https://ui.adsabs.harvard.edu/abs/2021TNSAN...7....1S}
}

@article{Smith2020,
	title        = {{Design and Operation of the ATLAS Transient Science Server}},
	author       = {{Smith}, K.~W. and {Smartt}, S.~J. and {Young}, D.~R. and {Tonry}, J.~L. and {Denneau}, L. and {Flewelling}, H. and {Heinze}, A.~N. and {Weiland}, H.~J. and {Stalder}, B. and {Rest}, A. and {Stubbs}, C.~W. and {Anderson}, J.~P. and {Chen}, T. -W. and {Clark}, P. and {Do}, A. and {F{\"o}rster}, F. and {Fulton}, M. and {Gillanders}, J. and {McBrien}, O.~R. and {O'Neill}, D. and {Srivastav}, S. and {Wright}, D.~E.},
	year         = 2020,
	month        = {aug},
	journal      = {\pasp},
	volume       = 132,
	number       = 1014,
	pages        = {085002},
	doi          = {10.1088/1538-3873/ab936e},
	eid          = {085002},
	archiveprefix = {arXiv},
	eprint       = {2003.09052},
	primaryclass = {astro-ph.IM},
	adsurl       = {https://ui.adsabs.harvard.edu/abs/2020PASP..132h5002S}
}

@article{pypeit:joss_pub,
	title        = {PypeIt: The Python Spectroscopic Data Reduction Pipeline},
	author       = {J. Xavier Prochaska and Joseph F. Hennawi and Kyle B. Westfall and Ryan J. Cooke and Feige Wang and Tiffany Hsyu and Frederick B. Davies and Emanuele Paolo Farina and Debora Pelliccia},
	year         = 2020,
	journal      = {Journal of Open Source Software},
	publisher    = {The Open Journal},
	volume       = 5,
	number       = 56,
	pages        = 2308,
	doi          = {10.21105/joss.02308},
	url          = {https://doi.org/10.21105/joss.02308}
}

@misc{pypeit:zenodo,
	title        = {{pypeit/PypeIt: Release 1.0.0}},
	author       = {{Prochaska}, J. Xavier and {Hennawi}, Joseph and {Cooke}, Ryan and {Westfall}, Kyle and {Wang}, Feige and {EmAstro} and {Tiffanyhsyu} and {Wasserman}, Asher and {Villaume}, Alexa and {Marijana777} and {Schindler}, JT and {Young}, David and {Simha}, Sunil and {Wilde}, Matt and {Tejos}, Nicolas and {Isbell}, Jacob and {Fl{\"o}rs}, Andreas and {Sandford}, Nathan and {Vasovi{\'c}}, Zlatan and {Betts}, Edward and {Holden}, Brad},
	year         = 2020,
	month        = {apr},
	publisher    = {Zenodo},
	doi          = {10.5281/zenodo.3743493},
	eid          = {10.5281/zenodo.3743493},
	version      = {v1.0.0},
	adsurl       = {https://ui.adsabs.harvard.edu/abs/2020zndo...3743493P}
}

@article{Puls2008,
	title        = {{Mass loss from hot massive stars}},
	author       = {{Puls}, Joachim and {Vink}, Jorick S. and {Najarro}, Francisco},
	year         = 2008,
	month        = {dec},
	journal      = {\aapr},
	volume       = 16,
	number       = {3-4},
	pages        = {209--325},
	doi          = {10.1007/s00159-008-0015-8},
	archiveprefix = {arXiv},
	eprint       = {0811.0487},
	primaryclass = {astro-ph},
	adsurl       = {https://ui.adsabs.harvard.edu/abs/2008A&ARv..16..209P}
}

@article{Marchant2024,
	title        = {{The Evolution of Massive Binary Stars}},
	author       = {{Marchant}, Pablo and {Bodensteiner}, Julia},
	year         = 2024,
	month        = {sep},
	journal      = {\araa},
	volume       = 62,
	number       = 1,
	pages        = {21--61},
	doi          = {10.1146/annurev-astro-052722-105936},
	archiveprefix = {arXiv},
	eprint       = {2311.01865},
	primaryclass = {astro-ph.SR},
	adsurl       = {https://ui.adsabs.harvard.edu/abs/2024ARA&A..62...21M}
}

@incollection{Gal-Yam2017,
	title        = {{Observational and Physical Classification of Supernovae}},
	author       = {{Gal-Yam}, Avishay},
	year         = 2017,
	booktitle    = {Handbook of Supernovae},
	publisher    = {Springer},
	pages        = 195,
	doi          = {10.1007/978-3-319-21846-5_35},
	editor       = {{Alsabti}, Athem W. and {Murdin}, Paul},
	adsurl       = {https://ui.adsabs.harvard.edu/abs/2017hsn..book..195G}
}

@article{Chen2018,
	title        = {{SN 2017ens: The Metamorphosis of a Luminous Broadlined Type Ic Supernova into an SN IIn}},
	author       = {{Chen}, T. -W. and {Inserra}, C. and {Fraser}, M. and {Moriya}, T.~J. and {Schady}, P. and {Schweyer}, T. and {Filippenko}, A.~V. and {Perley}, D.~A. and {Ruiter}, A.~J. and {Seitenzahl}, I. and {Sollerman}, J. and {Taddia}, F. and {Anderson}, J.~P. and {Foley}, R.~J. and {Jerkstrand}, A. and {Ngeow}, C. -C. and {Pan}, Y. -C. and {Pastorello}, A. and {Points}, S. and {Smartt}, S.~J. and {Smith}, K.~W. and {Taubenberger}, S. and {Wiseman}, P. and {Young}, D.~R. and {Benetti}, S. and {Berton}, M. and {Bufano}, F. and {Clark}, P. and {Della Valle}, M. and {Galbany}, L. and {Gal-Yam}, A. and {Gromadzki}, M. and {Guti{\'e}rrez}, C.~P. and {Heinze}, A. and {Kankare}, E. and {Kilpatrick}, C.~D. and {Kuncarayakti}, H. and {Leloudas}, G. and {Lin}, Z. -Y. and {Maguire}, K. and {Mazzali}, P. and {McBrien}, O. and {Prentice}, S.~J. and {Rau}, A. and {Rest}, A. and {Siebert}, M.~R. and {Stalder}, B. and {Tonry}, J.~L. and {Yu}, P. -C.},
	year         = 2018,
	month        = {nov},
	journal      = {\apjl},
	volume       = 867,
	number       = 2,
	pages        = {L31},
	doi          = {10.3847/2041-8213/aaeb2e},
	eid          = {L31},
	archiveprefix = {arXiv},
	eprint       = {1808.04382},
	primaryclass = {astro-ph.SR},
	adsurl       = {https://ui.adsabs.harvard.edu/abs/2018ApJ...867L..31C}
}

@article{Tonry2018,
	title        = {{The ATLAS All-Sky Stellar Reference Catalog}},
	author       = {{Tonry}, J.~L. and {Denneau}, L. and {Flewelling}, H. and {Heinze}, A.~N. and {Onken}, C.~A. and {Smartt}, S.~J. and {Stalder}, B. and {Weiland}, H.~J. and {Wolf}, C.},
	year         = 2018,
	month        = {nov},
	journal      = {\apj},
	volume       = 867,
	number       = 2,
	pages        = 105,
	doi          = {10.3847/1538-4357/aae386},
	eid          = 105,
	archiveprefix = {arXiv},
	eprint       = {1809.09157},
	primaryclass = {astro-ph.IM},
	adsurl       = {https://ui.adsabs.harvard.edu/abs/2018ApJ...867..105T}
}

@article{Fox2014,
	title        = {{Uncovering the Putative B-star Binary Companion of the SN 1993J Progenitor}},
	author       = {{Fox}, Ori D. and {Azalee Bostroem}, K. and {Van Dyk}, Schuyler D. and {Filippenko}, Alexei V. and {Fransson}, Claes and {Matheson}, Thomas and {Cenko}, S. Bradley and {Chandra}, Poonam and {Dwarkadas}, Vikram and {Li}, Weidong and {Parker}, Alex H. and {Smith}, Nathan},
	year         = 2014,
	month        = {jul},
	journal      = {\apj},
	volume       = 790,
	number       = 1,
	pages        = 17,
	doi          = {10.1088/0004-637X/790/1/17},
	eid          = 17,
	archiveprefix = {arXiv},
	eprint       = {1405.4863},
	primaryclass = {astro-ph.HE},
	adsurl       = {https://ui.adsabs.harvard.edu/abs/2014ApJ...790...17F}
}

@article{vanBaal2024,
	title        = {{Diagnostics of 3D explosion asymmetries of stripped-envelope supernovae by nebular line profiles}},
	author       = {{van Baal}, Bart F.~A. and {Jerkstrand}, Anders and {Wongwathanarat}, Annop and {Janka}, Hans-Thomas},
	year         = 2024,
	month        = {aug},
	journal      = {\mnras},
	volume       = 532,
	number       = 4,
	pages        = {4106--4131},
	doi          = {10.1093/mnras/stae1603},
	archiveprefix = {arXiv},
	eprint       = {2404.01763},
	primaryclass = {astro-ph.HE},
	adsurl       = {https://ui.adsabs.harvard.edu/abs/2024MNRAS.532.4106V}
}

@article{Masci2019,
	title        = {{The Zwicky Transient Facility: Data Processing, Products, and Archive}},
	author       = {{Masci}, Frank J. and {Laher}, Russ R. and {Rusholme}, Ben and {Shupe}, David L. and {Groom}, Steven and {Surace}, Jason and {Jackson}, Edward and {Monkewitz}, Serge and {Beck}, Ron and {Flynn}, David and {Terek}, Scott and {Landry}, Walter and {Hacopians}, Eugean and {Desai}, Vandana and {Howell}, Justin and {Brooke}, Tim and {Imel}, David and {Wachter}, Stefanie and {Ye}, Quan-Zhi and {Lin}, Hsing-Wen and {Cenko}, S. Bradley and {Cunningham}, Virginia and {Rebbapragada}, Umaa and {Bue}, Brian and {Miller}, Adam A. and {Mahabal}, Ashish and {Bellm}, Eric C. and {Patterson}, Maria T. and {Juri{\'c}}, Mario and {Golkhou}, V. Zach and {Ofek}, Eran O. and {Walters}, Richard and {Graham}, Matthew and {Kasliwal}, Mansi M. and {Dekany}, Richard G. and {Kupfer}, Thomas and {Burdge}, Kevin and {Cannella}, Christopher B. and {Barlow}, Tom and {Van Sistine}, Angela and {Giomi}, Matteo and {Fremling}, Christoffer and {Blagorodnova}, Nadejda and {Levitan}, David and {Riddle}, Reed and {Smith}, Roger M. and {Helou}, George and {Prince}, Thomas A. and {Kulkarni}, Shrinivas R.},
	year         = 2019,
	month        = {jan},
	journal      = {\pasp},
	volume       = 131,
	number       = 995,
	pages        = {018003},
	doi          = {10.1088/1538-3873/aae8ac},
	archiveprefix = {arXiv},
	eprint       = {1902.01872},
	primaryclass = {astro-ph.IM},
	adsurl       = {https://ui.adsabs.harvard.edu/abs/2019PASP..131a8003M}
}

@article{Dekany2020,
	title        = {{The Zwicky Transient Facility: Observing System}},
	author       = {{Dekany}, Richard and {Smith}, Roger M. and {Riddle}, Reed and {Feeney}, Michael and {Porter}, Michael and {Hale}, David and {Zolkower}, Jeffry and {Belicki}, Justin and {Kaye}, Stephen and {Henning}, John and {Walters}, Richard and {Cromer}, John and {Delacroix}, Alex and {Rodriguez}, Hector and {Reiley}, Daniel J. and {Mao}, Peter and {Hover}, David and {Murphy}, Patrick and {Burruss}, Rick and {Baker}, John and {Kowalski}, Marek and {Reif}, Klaus and {Mueller}, Phillip and {Bellm}, Eric and {Graham}, Matthew and {Kulkarni}, Shrinivas R.},
	year         = 2020,
	month        = {mar},
	journal      = {\pasp},
	volume       = 132,
	number       = 1009,
	pages        = {038001},
	doi          = {10.1088/1538-3873/ab4ca2},
	eid          = {038001},
	archiveprefix = {arXiv},
	eprint       = {2008.04923},
	primaryclass = {astro-ph.IM},
	adsurl       = {https://ui.adsabs.harvard.edu/abs/2020PASP..132c8001D}
}

@article{Graham2019,
	title        = {{The Zwicky Transient Facility: Science Objectives}},
	author       = {{Graham}, Matthew J. and {Kulkarni}, S.~R. and {Bellm}, Eric C. and {Adams}, Scott M. and {Barbarino}, Cristina and {Blagorodnova}, Nadejda and {Bodewits}, Dennis and {Bolin}, Bryce and {Brady}, Patrick R. and {Cenko}, S. Bradley and {Chang}, Chan-Kao and {Coughlin}, Michael W. and {De}, Kishalay and {Eadie}, Gwendolyn and {Farnham}, Tony L. and {Feindt}, Ulrich and {Franckowiak}, Anna and {Fremling}, Christoffer and {Gezari}, Suvi and {Ghosh}, Shaon and {Goldstein}, Daniel A. and {Golkhou}, V. Zach and {Goobar}, Ariel and {Ho}, Anna Y.~Q. and {Huppenkothen}, Daniela and {Ivezi{\'c}}, {\v{Z}}eljko and {Jones}, R. Lynne and {Juric}, Mario and {Kaplan}, David L. and {Kasliwal}, Mansi M. and {Kelley}, Michael S.~P. and {Kupfer}, Thomas and {Lee}, Chien-De and {Lin}, Hsing Wen and {Lunnan}, Ragnhild and {Mahabal}, Ashish A. and {Miller}, Adam A. and {Ngeow}, Chow-Choong and {Nugent}, Peter and {Ofek}, Eran O. and {Prince}, Thomas A. and {Rauch}, Ludwig and {van Roestel}, Jan and {Schulze}, Steve and {Singer}, Leo P. and {Sollerman}, Jesper and {Taddia}, Francesco and {Yan}, Lin and {Ye}, Quan-Zhi and {Yu}, Po-Chieh and {Barlow}, Tom and {Bauer}, James and {Beck}, Ron and {Belicki}, Justin and {Biswas}, Rahul and {Brinnel}, Valery and {Brooke}, Tim and {Bue}, Brian and {Bulla}, Mattia and {Burruss}, Rick and {Connolly}, Andrew and {Cromer}, John and {Cunningham}, Virginia and {Dekany}, Richard and {Delacroix}, Alex and {Desai}, Vandana and {Duev}, Dmitry A. and {Feeney}, Michael and {Flynn}, David and {Frederick}, Sara and {Gal-Yam}, Avishay and {Giomi}, Matteo and {Groom}, Steven and {Hacopians}, Eugean and {Hale}, David and {Helou}, George and {Henning}, John and {Hover}, David and {Hillenbrand}, Lynne A. and {Howell}, Justin and {Hung}, Tiara and {Imel}, David and {Ip}, Wing-Huen and {Jackson}, Edward and {Kaspi}, Shai and {Kaye}, Stephen and {Kowalski}, Marek and {Kramer}, Emily and {Kuhn}, Michael and {Landry}, Walter and {Laher}, Russ R. and {Mao}, Peter and {Masci}, Frank J. and {Monkewitz}, Serge and {Murphy}, Patrick and {Nordin}, Jakob and {Patterson}, Maria T. and {Penprase}, Bryan and {Porter}, Michael and {Rebbapragada}, Umaa and {Reiley}, Dan and {Riddle}, Reed and {Rigault}, Mickael and {Rodriguez}, Hector and {Rusholme}, Ben and {van Santen}, Jakob and {Shupe}, David L. and {Smith}, Roger M. and {Soumagnac}, Maayane T. and {Stein}, Robert and {Surace}, Jason and {Szkody}, Paula and {Terek}, Scott and {Van Sistine}, Angela and {van Velzen}, Sjoert and {Vestrand}, W. Thomas and {Walters}, Richard and {Ward}, Charlotte and {Zhang}, Chaoran and {Zolkower}, Jeffry},
	year         = 2019,
	month        = {jul},
	journal      = {\pasp},
	volume       = 131,
	number       = 1001,
	pages        = {078001},
	doi          = {10.1088/1538-3873/ab006c},
	archiveprefix = {arXiv},
	eprint       = {1902.01945},
	primaryclass = {astro-ph.IM},
	adsurl       = {https://ui.adsabs.harvard.edu/abs/2019PASP..131g8001G}
}

@article{Bellm2019,
	title        = {{The Zwicky Transient Facility: System Overview, Performance, and First Results}},
	author       = {{Bellm}, Eric C. and {Kulkarni}, Shrinivas R. and {Graham}, Matthew J. and {Dekany}, Richard and {Smith}, Roger M. and {Riddle}, Reed and {Masci}, Frank J. and {Helou}, George and {Prince}, Thomas A. and {Adams}, Scott M. and {Barbarino}, C. and {Barlow}, Tom and {Bauer}, James and {Beck}, Ron and {Belicki}, Justin and {Biswas}, Rahul and {Blagorodnova}, Nadejda and {Bodewits}, Dennis and {Bolin}, Bryce and {Brinnel}, Valery and {Brooke}, Tim and {Bue}, Brian and {Bulla}, Mattia and {Burruss}, Rick and {Cenko}, S. Bradley and {Chang}, Chan-Kao and {Connolly}, Andrew and {Coughlin}, Michael and {Cromer}, John and {Cunningham}, Virginia and {De}, Kishalay and {Delacroix}, Alex and {Desai}, Vandana and {Duev}, Dmitry A. and {Eadie}, Gwendolyn and {Farnham}, Tony L. and {Feeney}, Michael and {Feindt}, Ulrich and {Flynn}, David and {Franckowiak}, Anna and {Frederick}, S. and {Fremling}, C. and {Gal-Yam}, Avishay and {Gezari}, Suvi and {Giomi}, Matteo and {Goldstein}, Daniel A. and {Golkhou}, V. Zach and {Goobar}, Ariel and {Groom}, Steven and {Hacopians}, Eugean and {Hale}, David and {Henning}, John and {Ho}, Anna Y.~Q. and {Hover}, David and {Howell}, Justin and {Hung}, Tiara and {Huppenkothen}, Daniela and {Imel}, David and {Ip}, Wing-Huen and {Ivezi{\'c}}, {\v{Z}}eljko and {Jackson}, Edward and {Jones}, Lynne and {Juric}, Mario and {Kasliwal}, Mansi M. and {Kaspi}, S. and {Kaye}, Stephen and {Kelley}, Michael S.~P. and {Kowalski}, Marek and {Kramer}, Emily and {Kupfer}, Thomas and {Landry}, Walter and {Laher}, Russ R. and {Lee}, Chien-De and {Lin}, Hsing Wen and {Lin}, Zhong-Yi and {Lunnan}, Ragnhild and {Giomi}, Matteo and {Mahabal}, Ashish and {Mao}, Peter and {Miller}, Adam A. and {Monkewitz}, Serge and {Murphy}, Patrick and {Ngeow}, Chow-Choong and {Nordin}, Jakob and {Nugent}, Peter and {Ofek}, Eran and {Patterson}, Maria T. and {Penprase}, Bryan and {Porter}, Michael and {Rauch}, Ludwig and {Rebbapragada}, Umaa and {Reiley}, Dan and {Rigault}, Mickael and {Rodriguez}, Hector and {van Roestel}, Jan and {Rusholme}, Ben and {van Santen}, Jakob and {Schulze}, S. and {Shupe}, David L. and {Singer}, Leo P. and {Soumagnac}, Maayane T. and {Stein}, Robert and {Surace}, Jason and {Sollerman}, Jesper and {Szkody}, Paula and {Taddia}, F. and {Terek}, Scott and {Van Sistine}, Angela and {van Velzen}, Sjoert and {Vestrand}, W. Thomas and {Walters}, Richard and {Ward}, Charlotte and {Ye}, Quan-Zhi and {Yu}, Po-Chieh and {Yan}, Lin and {Zolkower}, Jeffry},
	year         = 2019,
	month        = {jan},
	journal      = {\pasp},
	volume       = 131,
	number       = 995,
	pages        = {018002},
	doi          = {10.1088/1538-3873/aaecbe},
	archiveprefix = {arXiv},
	eprint       = {1902.01932},
	primaryclass = {astro-ph.IM},
	adsurl       = {https://ui.adsabs.harvard.edu/abs/2019PASP..131a8002B}
}

@article{Matheson2000,
	title        = {{Detailed Analysis of Early to Late-Time Spectra of Supernova 1993J}},
	author       = {{Matheson}, Thomas and {Filippenko}, Alexei V. and {Ho}, Luis C. and {Barth}, Aaron J. and {Leonard}, Douglas C.},
	year         = 2000,
	month        = {sep},
	journal      = {\aj},
	volume       = 120,
	number       = 3,
	pages        = {1499--1515},
	doi          = {10.1086/301519},
	archiveprefix = {arXiv},
	eprint       = {astro-ph/0006264},
	primaryclass = {astro-ph},
	adsurl       = {https://ui.adsabs.harvard.edu/abs/2000AJ....120.1499M}
}

@article{Filippenko1997,
	title        = {{Optical Spectra of Supernovae}},
	author       = {{Filippenko}, Alexei V.},
	year         = 1997,
	month        = {jan},
	journal      = {\araa},
	volume       = 35,
	pages        = {309--355},
	doi          = {10.1146/annurev.astro.35.1.309},
	adsurl       = {https://ui.adsabs.harvard.edu/abs/1997ARA&A..35..309F}
}

@incollection{Jerkstrand2017,
	title        = {{Spectra of Supernovae in the Nebular Phase}},
	author       = {{Jerkstrand}, Anders},
	year         = 2017,
	booktitle    = {Handbook of Supernovae},
	pages        = 795,
	doi          = {10.1007/978-3-319-21846-5_29},
	editor       = {{Alsabti}, Athem W. and {Murdin}, Paul},
	adsurl       = {https://ui.adsabs.harvard.edu/abs/2017hsn..book..795J}
}

@article{Woosley2019,
	title        = {{The Evolution of Massive Helium Stars, Including Mass Loss}},
	author       = {{Woosley}, S.~E.},
	year         = 2019,
	month        = {jun},
	journal      = {\apj},
	volume       = 878,
	number       = 1,
	pages        = 49,
	doi          = {10.3847/1538-4357/ab1b41},
	eid          = 49,
	archiveprefix = {arXiv},
	eprint       = {1901.00215},
	primaryclass = {astro-ph.SR},
	adsurl       = {https://ui.adsabs.harvard.edu/abs/2019ApJ...878...49W}
}

@article{Ertl2020,
	title        = {{The Explosion of Helium Stars Evolved with Mass Loss}},
	author       = {{Ertl}, T. and {Woosley}, S.~E. and {Sukhbold}, Tuguldur and {Janka}, H. -T.},
	year         = 2020,
	month        = {feb},
	journal      = {\apj},
	volume       = 890,
	number       = 1,
	pages        = 51,
	doi          = {10.3847/1538-4357/ab6458},
	eid          = 51,
	archiveprefix = {arXiv},
	eprint       = {1910.01641},
	primaryclass = {astro-ph.HE},
	adsurl       = {https://ui.adsabs.harvard.edu/abs/2020ApJ...890...51E}
}

@article{Jerkstrand2015,
	title        = {{Late-time spectral line formation in Type IIb supernovae, with application to SN 1993J, SN 2008ax, and SN 2011dh}},
	author       = {{Jerkstrand}, A. and {Ergon}, M. and {Smartt}, S.~J. and {Fransson}, C. and {Sollerman}, J. and {Taubenberger}, S. and {Bersten}, M. and {Spyromilio}, J.},
	year         = 2015,
	month        = {jan},
	journal      = {\aap},
	volume       = 573,
	pages        = {A12},
	doi          = {10.1051/0004-6361/201423983},
	eid          = {A12},
	archiveprefix = {arXiv},
	eprint       = {1408.0732},
	primaryclass = {astro-ph.HE},
	adsurl       = {https://ui.adsabs.harvard.edu/abs/2015A&A...573A..12J}
}

@phdthesis{Jerkstrand2011,
	title        = {{Spectral modeling of nebular-phase supernovae}},
	author       = {{Jerkstrand}, Anders},
	year         = 2011,
	month        = {dec},
	school       = {Stockholm University},
	adsurl       = {https://ui.adsabs.harvard.edu/abs/2011PhDT........90J}
}

@article{Jerkstrand2012,
	title        = {{The progenitor mass of the Type IIP supernova SN 2004et from late-time spectral modeling}},
	author       = {{Jerkstrand}, A. and {Fransson}, C. and {Maguire}, K. and {Smartt}, S. and {Ergon}, M. and {Spyromilio}, J.},
	year         = 2012,
	month        = {oct},
	journal      = {\aap},
	volume       = 546,
	pages        = {A28},
	doi          = {10.1051/0004-6361/201219528},
	eid          = {A28},
	archiveprefix = {arXiv},
	eprint       = {1208.2183},
	primaryclass = {astro-ph.HE},
	adsurl       = {https://ui.adsabs.harvard.edu/abs/2012A&A...546A..28J}
}

@article{Jerkstrand2014,
	title        = {{The nebular spectra of SN 2012aw and constraints on stellar nucleosynthesis from oxygen emission lines}},
	author       = {{Jerkstrand}, A. and {Smartt}, S.~J. and {Fraser}, M. and {Fransson}, C. and {Sollerman}, J. and {Taddia}, F. and {Kotak}, R.},
	year         = 2014,
	month        = {apr},
	journal      = {\mnras},
	volume       = 439,
	number       = 4,
	pages        = {3694--3703},
	doi          = {10.1093/mnras/stu221},
	archiveprefix = {arXiv},
	eprint       = {1311.2031},
	primaryclass = {astro-ph.SR},
	adsurl       = {https://ui.adsabs.harvard.edu/abs/2014MNRAS.439.3694J}
}

@article{Dessart2023,
	title        = {{Modeling of the nebular-phase spectral evolution of stripped-envelope supernovae. New grids from 100 to 450 days}},
	author       = {{Dessart}, L. and {Hillier}, D. John and {Woosley}, S.~E. and {Kuncarayakti}, H.},
	year         = 2023,
	month        = {sep},
	journal      = {\aap},
	volume       = 677,
	pages        = {A7},
	doi          = {10.1051/0004-6361/202346626},
	eid          = {A7},
	archiveprefix = {arXiv},
	eprint       = {2306.12092},
	primaryclass = {astro-ph.SR},
	adsurl       = {https://ui.adsabs.harvard.edu/abs/2023A&A...677A...7D}
}

@ARTICLE{Smith2024,
       author = {{Smith}, K.~W. and {Young}, D.~R. and {Nicholl}, M. and {Fulton}, M. and {McCollum}, M. and {Moore}, T. and {Weston}, J. and {Sheng}, X. and {Aamer}, A. and {Angus}, C.~R. and {Ramsden}, P. and {Shingles}, L. and {Smartt}, S.~J. and {Srivastav}, S. and {Gillanders}, J. and {Rhodes}, L. and {Andersson}, A. and {Stevance}, H. and {Denneau}, L. and {Tonry}, J. and {Weiland}, H. and {Lawrence}, A. and {Siverd}, R. and {Erasmus}, N. and {Koorts}, W. and {Jordan}, A. and {Suc}, V. and {Rest}, A. and {Chen}, T.~W. and {Stubbs}, C. and {Sommer}, J.},
        title = "{ATLAS24dqq (AT2024ehs): discovery of a rapidly brightening candidate SN in NGC 3443 (25 Mpc)}",
      journal = {Transient Name Server AstroNote},
         year = 2024,
        month = mar,
       volume = {78},
        pages = {1},
       adsurl = {https://ui.adsabs.harvard.edu/abs/2024TNSAN..78....1S}
}

@ARTICLE{Tonry2024,
       author = {{Tonry}, J. and {Denneau}, L. and {Weiland}, H. and {Lawrence}, A. and {Siverd}, R. and {Erasmus}, N. and {Koorts}, W. and {Jordan}, A. and {Suc}, V. and {Smartt}, S.~J. and {Smith}, K.~W. and {Young}, D.~R. and {Nicholl}, M. and {Fulton}, M. and {McCollum}, M. and {Moore}, T. and {Weston}, J. and {Sheng}, X. and {Ramsden}, P. and {Angus}, C.~R. and {Aamer}, A. and {Shingles}, L. and {Srivastav}, S. and {Gillanders}, J. and {Rhodes}, L. and {Andersson}, A. and {Stevance}, H. and {Rest}, A. and {Chen}, T.~W. and {Stubbs}, C. and {Sommer}, J.},
        title = "{ATLAS Transient Discovery Report for 2024-03-15}",
      journal = {Transient Name Server Discovery Report},
         year = 2024,
        month = mar,
       volume = {2024-701},
        pages = {1},
       adsurl = {https://ui.adsabs.harvard.edu/abs/2024TNSTR.701....1T}
}

@ARTICLE{Filippenko1988,
       author = {{Filippenko}, Alexei V.},
        title = "{Supernova 1987K: Type II in Youth, Type Ib in Old Age}",
      journal = {\aj},
         year = 1988,
        month = dec,
       volume = {96},
        pages = {1941},
          doi = {10.1086/114940},
       adsurl = {https://ui.adsabs.harvard.edu/abs/1988AJ.....96.1941F}
}

@book{BranchWheeler,
    author = {{Branch}, David and {Wheeler}, J.},
    year = {2017},
    month = {01},
    pages = {},
    publisher = {Springer Berlin, Heidelberg},
    title = {Supernova Explosions},
    isbn = {978-3-662-55052-6},
    doi = {10.1007/978-3-662-55054-0}
}

@ARTICLE{Sana2012,
       author = {{Sana}, H. and {de Mink}, S.~E. and {de Koter}, A. and {Langer}, N. and {Evans}, C.~J. and {Gieles}, M. and {Gosset}, E. and {Izzard}, R.~G. and {Le Bouquin}, J. -B. and {Schneider}, F.~R.~N.},
        title = "{Binary Interaction Dominates the Evolution of Massive Stars}",
      journal = {Science},
         year = 2012,
        month = jul,
       volume = {337},
       number = {6093},
        pages = {444},
          doi = {10.1126/science.1223344},
archivePrefix = {arXiv},
       eprint = {1207.6397},
 primaryClass = {astro-ph.SR},
       adsurl = {https://ui.adsabs.harvard.edu/abs/2012Sci...337..444S}
}

@ARTICLE{Gilkis2022,
       author = {{Gilkis}, Avishai and {Arcavi}, Iair},
        title = "{How much hydrogen is in Type Ib and IIb supernova progenitors?}",
      journal = {\mnras},
         year = 2022,
        month = mar,
       volume = {511},
       number = {1},
        pages = {691-712},
          doi = {10.1093/mnras/stac088},
archivePrefix = {arXiv},
       eprint = {2111.04432},
 primaryClass = {astro-ph.SR},
       adsurl = {https://ui.adsabs.harvard.edu/abs/2022MNRAS.511..691G}
}

@article{Stritzinger2009,
   title={THE HE-RICH CORE-COLLAPSE SUPERNOVA 2007Y: OBSERVATIONS FROM X-RAY TO RADIO WAVELENGTHS},
   volume={696},
   ISSN={1538-4357},
   url={http://dx.doi.org/10.1088/0004-637X/696/1/713},
   DOI={10.1088/0004-637x/696/1/713},
   number={1},
   journal={The Astrophysical Journal},
   publisher={American Astronomical Society},
   author={Stritzinger, Maximilian and Mazzali, Paolo and Phillips, Mark M. and Immler, Stefan and Soderberg, Alicia and Sollerman, Jesper and Boldt, Luis and Braithwaite, Jonathan and Brown, Peter and Burns, Christopher R. and Contreras, Carlos and Covarrubias, Ricardo and Folatelli, Gastón and Freedman, Wendy L. and González, Sergio and Hamuy, Mario and Krzeminski, Wojtek and Madore, Barry F. and Milne, Peter and Morrell, Nidia and Persson, S. E. and Roth, Miguel and Smith, Mathew and Suntzeff, Nicholas B.},
   year={2009},
   month=apr, pages={713–728} }

@ARTICLE{Valenti2011,
  author={Valenti, S. and Fraser, M. and Benetti, S. and Pignata, G. and Sollerman, J. and Inserra, C. and Cappellaro, E. and Pastorello, A. and Smartt, S. J. and Ergon, M. and Botticella, M. T. and Brimacombe, J. and Bufano, F. and Crockett, M. and Eder, I. and Fugazza, D. and Haislip, J. B and Hamuy, M. and Harutyunyan, A. and Ivarsen, K. M. and Kankare, E. and Kotak, R. and LaCluyze, A. P. and Magill, L. and Mattila, S. and Maza, J. and Mazzali, P. A. and Reichart, D. E. and Taubenberger, S. and Turatto, M. and Zampieri, L.},
  journal={Monthly Notices of the Royal Astronomical Society}, 
  title={SN 2009jf: a slow-evolving stripped-envelope core-collapse supernova}, 
  year={2011},
  volume={416},
  number={4},
  pages={3138-3159},
  doi={10.1111/j.1365-2966.2011.19262.x}}

@article{Kumar2013,
   title={Light curve and spectral evolution of the Type IIb supernova 2011fu},
   volume={431},
   ISSN={1365-2966},
   url={http://dx.doi.org/10.1093/mnras/stt162},
   DOI={10.1093/mnras/stt162},
   number={1},
   journal={Monthly Notices of the Royal Astronomical Society},
   publisher={Oxford University Press (OUP)},
   author={Kumar, Brajesh and Pandey, S. B. and Sahu, D. K. and Vinko, J. and Moskvitin, A. S. and Anupama, G. C. and Bhatt, V. K. and Ordasi, A. and Nagy, A. and Sokolov, V. V. and Sokolova, T. N. and Komarova, V. N. and Kumar, Brijesh and Bose, Subhash and Roy, Rupak and Sagar, Ram},
   year={2013},
   month=feb, pages={308–321} }

@article{MoralesGaroffolo2015,
   title={SN 2011fu: a type IIb supernova with a luminous double-peaked light curve},
   volume={454},
   ISSN={1365-2966},
   url={http://dx.doi.org/10.1093/mnras/stv1972},
   DOI={10.1093/mnras/stv1972},
   number={1},
   journal={Monthly Notices of the Royal Astronomical Society},
   publisher={Oxford University Press (OUP)},
   author={Morales-Garoffolo, A. and Elias-Rosa, N. and Bersten, M. and Jerkstrand, A. and Taubenberger, S. and Benetti, S. and Cappellaro, E. and Kotak, R. and Pastorello, A. and Bufano, F. and Domínguez, R. M. and Ergon, M. and Fraser, M. and Gao, X. and García, E. and Howell, D. A. and Isern, J. and Smartt, S. J. and Tomasella, L. and Valenti, S.},
   year={2015},
   month=sep, pages={95–114} }

@ARTICLE{Pandey2021,
       author = {{Pandey}, S.~B. and {Kumar}, Amit and {Kumar}, Brajesh and {Anupama}, G.~C. and {Srivastav}, S. and {Sahu}, D.~K. and {Vinko}, J. and {Aryan}, A. and {Pastorello}, A. and {Benetti}, S. and {Tomasella}, L. and {Singh}, Avinash and {Moskvitin}, A.~S. and {Sokolov}, V.~V. and {Gupta}, R. and {Misra}, K. and {Ochner}, P. and {Valenti}, S.},
        title = "{Photometric, polarimetric, and spectroscopic studies of the luminous, slow-decaying Type Ib SN 2012au}",
      journal = {\mnras},
         year = 2021,
        month = oct,
       volume = {507},
       number = {1},
        pages = {1229-1253},
          doi = {10.1093/mnras/stab1889},
archivePrefix = {arXiv},
       eprint = {2106.15856},
 primaryClass = {astro-ph.HE},
       adsurl = {https://ui.adsabs.harvard.edu/abs/2021MNRAS.507.1229P}
}

@article{Liu2016,
doi = {10.3847/0004-637X/827/2/90},
url = {https://dx.doi.org/10.3847/0004-637X/827/2/90},
year = {2016},
month = {aug},
publisher = {The American Astronomical Society},
volume = {827},
number = {2},
pages = {90},
author = {Yu-Qian Liu and Maryam Modjaz and Federica B. Bianco and Or Graur},
title = {ANALYZING THE LARGEST SPECTROSCOPIC DATA SET OF STRIPPED SUPERNOVAE TO IMPROVE THEIR IDENTIFICATIONS AND CONSTRAIN THEIR PROGENITORS},
journal = {The Astrophysical Journal}
}

@article{Maund2004,
   title={The massive binary companion star to the progenitor of supernova 1993J},
   volume={427},
   ISSN={1476-4687},
   url={http://dx.doi.org/10.1038/nature02161},
   DOI={10.1038/nature02161},
   number={6970},
   journal={Nature},
   publisher={Springer Science and Business Media LLC},
   author={Maund, Justyn R. and Smartt, Stephen J. and Kudritzki, Rolf P. and Podsiadlowski, Philipp and Gilmore, Gerard F.},
   year={2004},
   month=jan, pages={129–131} }

@article{Dessart2011,
    author = {Dessart, Luc and Hillier, D. John and Livne, Eli and Yoon, Sung-Chul and Woosley, Stan and Waldman, Roni and Langer, Norbert},
    title = {Core-collapse explosions of Wolf–Rayet stars and the connection to Type IIb/Ib/Ic supernovae},
    journal = {Monthly Notices of the Royal Astronomical Society},
    volume = {414},
    number = {4},
    pages = {2985-3005},
    year = {2011},
    month = {07},
    issn = {0035-8711},
    doi = {10.1111/j.1365-2966.2011.18598.x},
    url = {https://doi.org/10.1111/j.1365-2966.2011.18598.x},
    eprint = {https://academic.oup.com/mnras/article-pdf/414/4/2985/18701232/mnras0414-2985.pdf},
}

@ARTICLE{Shenar2024,
       author = {{Shenar}, Tomer},
        title = "{Wolf-Rayet stars}",
      journal = {arXiv e-prints},
         year = 2024,
        month = oct,
          eid = {arXiv:2410.04436},
        pages = {arXiv:2410.04436},
          doi = {10.48550/arXiv.2410.04436},
archivePrefix = {arXiv},
       eprint = {2410.04436},
 primaryClass = {astro-ph.SR},
       adsurl = {https://ui.adsabs.harvard.edu/abs/2024arXiv241004436S}
}

@ARTICLE{Krticka2014,
       author = {{Krti{\v{c}}ka}, J. and {Kub{\'a}t}, J.},
        title = "{Effect of rotational mixing and metallicity on the hot star wind mass-loss rates}",
      journal = {\aap},
         year = 2014,
        month = jul,
       volume = {567},
          eid = {A63},
        pages = {A63},
          doi = {10.1051/0004-6361/201423845},
archivePrefix = {arXiv},
       eprint = {1406.1288},
 primaryClass = {astro-ph.SR},
       adsurl = {https://ui.adsabs.harvard.edu/abs/2014A&A...567A..63K}
}

@ARTICLE{Gangopadhyay2023,
       author = {{Gangopadhyay}, Anjasha and {Maeda}, Keiichi and {Singh}, Avinash and {Nayana}, A.~J. and {Nakaoka}, Tatsuya and {Kawabata}, Koji S. and {Taguchi}, Kenta and {Singh}, Mridweeka and {Chandra}, Poonam and {Ryder}, Stuart D. and {Dastidar}, Raya and {Yamanaka}, Masayuki and {Kawabata}, Miho and {Alsaberi}, Rami Z.~E. and {Dukiya}, Naveen and {Teja}, Rishabh Singh and {Ailawadhi}, Bhavya and {Dutta}, Anirban and {Sahu}, D.~K. and {Moriya}, Takashi J. and {Misra}, Kuntal and {Tanaka}, Masaomi and {Chevalier}, Roger and {Tominaga}, Nozomu and {Uno}, Kohki and {Imazawa}, Ryo and {Hamada}, Taisei and {Hori}, Tomoya and {Isogai}, Keisuke},
        title = "{Bridging between Type IIb and Ib Supernovae: SN IIb 2022crv with a Very Thin Hydrogen Envelope}",
      journal = {\apj},
         year = 2023,
        month = nov,
       volume = {957},
       number = {2},
          eid = {100},
        pages = {100},
          doi = {10.3847/1538-4357/acfa94},
archivePrefix = {arXiv},
       eprint = {2309.07463},
 primaryClass = {astro-ph.HE},
       adsurl = {https://ui.adsabs.harvard.edu/abs/2023ApJ...957..100G}
}

@ARTICLE{Yoon2017,
       author = {{Yoon}, Sung-Chul and {Dessart}, Luc and {Clocchiatti}, Alejandro},
        title = "{Type Ib and IIb Supernova Progenitors in Interacting Binary Systems}",
      journal = {\apj},
         year = 2017,
        month = may,
       volume = {840},
       number = {1},
          eid = {10},
        pages = {10},
          doi = {10.3847/1538-4357/aa6afe},
archivePrefix = {arXiv},
       eprint = {1701.02089},
 primaryClass = {astro-ph.SR},
       adsurl = {https://ui.adsabs.harvard.edu/abs/2017ApJ...840...10Y}
}

@ARTICLE{Baade1934,
       author = {{Baade}, W. and {Zwicky}, F.},
        title = "{On Super-novae}",
      journal = {Proceedings of the National Academy of Science},
         year = 1934,
        month = may,
       volume = {20},
       number = {5},
        pages = {254-259},
          doi = {10.1073/pnas.20.5.254},
       adsurl = {https://ui.adsabs.harvard.edu/abs/1934PNAS...20..254B}
}

@ARTICLE{Minkowski1941,
       author = {{Minkowski}, R.},
        title = "{Spectra of Supernovae}",
      journal = {\pasp},
         year = 1941,
        month = aug,
       volume = {53},
       number = {314},
        pages = {224},
          doi = {10.1086/125315},
       adsurl = {https://ui.adsabs.harvard.edu/abs/1941PASP...53..224M}
}

@INPROCEEDINGS{sanaEvans2011,
       author = {{Sana}, Hugues and {Evans}, Christopher J.},
        title = "{The multiplicity of massive stars}",
    booktitle = {Active OB Stars: Structure, Evolution, Mass Loss, and Critical Limits},
         year = 2011,
       editor = {{Neiner}, Coralie and {Wade}, Gregg and {Meynet}, Georges and {Peters}, Geraldine},
       series = {IAU Symposium},
       volume = {272},
        month = jul,
        pages = {474-485},
          doi = {10.1017/S1743921311011124},
archivePrefix = {arXiv},
       eprint = {1009.4197},
 primaryClass = {astro-ph.SR},
       adsurl = {https://ui.adsabs.harvard.edu/abs/2011IAUS..272..474S}
}

@article{Smith2011,
    author = {Smith, Nathan and Li, Weidong and Filippenko, Alexei V. and Chornock, Ryan},
    title = {Observed fractions of core-collapse supernova types and initial masses of their single and binary progenitor stars},
    journal = {Monthly Notices of the Royal Astronomical Society},
    volume = {412},
    number = {3},
    pages = {1522-1538},
    year = {2011},
    month = {04},
    issn = {0035-8711},
    doi = {10.1111/j.1365-2966.2011.17229.x},
    url = {https://doi.org/10.1111/j.1365-2966.2011.17229.x},
    eprint = {https://academic.oup.com/mnras/article-pdf/412/3/1522/3576609/mnras0412-1522.pdf},
}

@ARTICLE{Filippenko1993,
       author = {{Filippenko}, Alexei V. and {Matheson}, Thomas and {Ho}, Luis C.},
        title = "{The ``Type IIb'' Supernova 1993J in M81: A Close Relative of Type Ib Supernovae}",
      journal = {\apjl},
         year = 1993,
        month = oct,
       volume = {415},
        pages = {L103},
          doi = {10.1086/187043},
       adsurl = {https://ui.adsabs.harvard.edu/abs/1993ApJ...415L.103F}
}

@ARTICLE{falk1977,
       author = {{Falk}, Sydney W. and {Arnett}, W. David},
        title = "{Radiation Dynamics, Envelope Ejection, and Supernova Light Curves}",
      journal = {\apjs},
         year = 1977,
        month = apr,
       volume = {33},
        pages = {515},
          doi = {10.1086/190440},
       adsurl = {https://ui.adsabs.harvard.edu/abs/1977ApJS...33..515F}
}

@ARTICLE{Arnett1982,
       author = {{Arnett}, W.~D.},
        title = "{Type I supernovae. I - Analytic solutions for the early part of the light curve}",
      journal = {\apj},
         year = 1982,
        month = feb,
       volume = {253},
        pages = {785-797},
          doi = {10.1086/159681},
       adsurl = {https://ui.adsabs.harvard.edu/abs/1982ApJ...253..785A}
}

@misc{jerkstrand2025corecollapsesupernovae,
      title={Core-collapse supernovae}, 
      author={Anders Jerkstrand and Dan Milisavljevic and Bernhard Müller},
      year={2025},
      eprint={2503.01321},
      archivePrefix={arXiv},
      primaryClass={astro-ph.HE},
      url={https://arxiv.org/abs/2503.01321}, 
}

@ARTICLE{patat1994,
       author = {{Patat}, F. and {Barbon}, R. and {Cappellaro}, E. and {Turatto}, M.},
        title = "{Light curves of type II supernovae. II. The analysis.}",
      journal = {\aap},
         year = 1994,
        month = feb,
       volume = {282},
        pages = {731-741},
       adsurl = {https://ui.adsabs.harvard.edu/abs/1994A&A...282..731P}
}

@ARTICLE{Medler2022acat,
       author = {{Medler}, K. and {Mazzali}, P.~A. and {Teffs}, J. and {Ashall}, C. and {Anderson}, J.~P. and {Arcavi}, I. and {Benetti}, S. and {Bostroem}, K.~A. and {Burke}, J. and {Cai}, Y. -Z. and {Charalampopoulos}, P. and {Elias-Rosa}, N. and {Ergon}, M. and {Galbany}, L. and {Gromadzki}, M. and {Hiramatsu}, D. and {Howell}, D.~A. and {Inserra}, C. and {Lundqvist}, P. and {McCully}, C. and {M{\"u}ller-Bravo}, T. and {Newsome}, M. and {Nicholl}, M. and {Padilla Gonzalez}, E. and {Paraskeva}, E. and {Pastorello}, A. and {Pellegrino}, C. and {Pessi}, P.~J. and {Reguitti}, A. and {Reynolds}, T.~M. and {Roy}, R. and {Terreran}, G. and {Tomasella}, L. and {Young}, D.~R.},
        title = "{SN 2020acat: an energetic fast rising Type IIb supernova}",
      journal = {\mnras},
         year = 2022,
        month = jul,
       volume = {513},
       number = {4},
        pages = {5540-5558},
          doi = {10.1093/mnras/stac1192},
archivePrefix = {arXiv},
       eprint = {2201.06991},
 primaryClass = {astro-ph.HE},
       adsurl = {https://ui.adsabs.harvard.edu/abs/2022MNRAS.513.5540M}
}

@ARTICLE{barmentloo2024,
       author = {{Barmentloo}, Stan and {Jerkstrand}, Anders and {Iwamoto}, Koichi and {Hachisu}, Izumi and {Nomoto}, Ken'ichi and {Sollerman}, Jesper and {Woosley}, Stan},
        title = "{Nebular nitrogen line emission in stripped-envelope supernovae - a new progenitor mass diagnostic}",
      journal = {\mnras},
         year = 2024,
        month = sep,
       volume = {533},
       number = {2},
        pages = {1251-1280},
          doi = {10.1093/mnras/stae1811},
archivePrefix = {arXiv},
       eprint = {2403.08911},
 primaryClass = {astro-ph.HE},
       adsurl = {https://ui.adsabs.harvard.edu/abs/2024MNRAS.533.1251B}
}

@ARTICLE{Leadbeater2024,
       author = {{Leadbeater}, R.},
        title = "{Transient Classification Report for 2024-03-31}",
      journal = {Transient Name Server Classification Report},
         year = 2024,
        month = mar,
       volume = {2024-872},
        pages = {1},
       adsurl = {https://ui.adsabs.harvard.edu/abs/2024TNSCR.872....1L}
}

@ARTICLE{Li2024,
       author = {{Li}, L. and {Cai}, Y. and {Zhai}, Q. and {Zhang}, J. and {Wang}, X. and {Pastorello}, A. and {Reguitti}, A.},
        title = "{LiONS Transient Classification Report for 2024-03-17}",
      journal = {Transient Name Server Classification Report},
         year = 2024,
        month = mar,
       volume = {2024-724},
        pages = {1},
       adsurl = {https://ui.adsabs.harvard.edu/abs/2024TNSCR.724....1L}
}

@misc{ergon2023,
      title={Lightcurve and spectral modelling of the Type IIb SN 2020acat. Evidence for a strong Ni bubble effect on the diffusion time}, 
      author={Mattias Ergon and Peter Lundqvist and Claes Fransson and Hanindyo Kuncarayakti and Kaustav K. Das and Kishalay De and Lucia Ferrari and Christoffer Fremling and Kyle Medler and Keiichi Maeda and Andrea Pastorello and Jesper Sollerman and Maximilian D. Stritzinger},
      year={2023},
      eprint={2308.07158},
      archivePrefix={arXiv},
      primaryClass={astro-ph.HE},
      url={https://arxiv.org/abs/2308.07158}, 
}

@article{Blinnikov1998,
   title={A Comparative Modeling of Supernova 1993J},
   volume={496},
   ISSN={1538-4357},
   url={http://dx.doi.org/10.1086/305375},
   DOI={10.1086/305375},
   number={1},
   journal={The Astrophysical Journal},
   publisher={American Astronomical Society},
   author={Blinnikov, S. I. and Eastman, R. and Bartunov, O. S. and Popolitov, V. A. and Woosley, S. E.},
   year={1998},
   month=mar, pages={454–472} }

@ARTICLE{Roy2013,
       author = {{Roy}, Rupak and {Kumar}, Brijesh and {Maund}, Justyn R. and {Schady}, Patricia and {Olivares}, E. Felipe and {Malesani}, Daniele and {Leloudas}, Giorgos and {Nandi}, Sumana and {Tanvir}, Nial and {Milisavljevic}, Dan and {Hjorth}, Jens and {Misra}, Kuntal and {Kumar}, Brajesh and {Pandey}, S.~B. and {Sagar}, Ram and {Chandola}, H.~C.},
        title = "{SN 2007uy - metamorphosis of an aspheric Type Ib explosion}",
      journal = {\mnras},
         year = 2013,
        month = sep,
       volume = {434},
       number = {3},
        pages = {2032-2050},
          doi = {10.1093/mnras/stt1148},
archivePrefix = {arXiv},
       eprint = {1306.5389},
 primaryClass = {astro-ph.HE},
       adsurl = {https://ui.adsabs.harvard.edu/abs/2013MNRAS.434.2032R}
}

@ARTICLE{Kumar2022,
       author = {{Kumar}, Brajesh and {Singh}, Avinash and {Sahu}, D.~K. and {Anupama}, G.~C.},
        title = "{Investigating the Observational Properties of Type Ib Supernova SN 2017iro}",
      journal = {\apj},
         year = 2022,
        month = mar,
       volume = {927},
       number = {1},
          eid = {61},
        pages = {61},
          doi = {10.3847/1538-4357/ac4bb9},
archivePrefix = {arXiv},
       eprint = {2201.03260},
 primaryClass = {astro-ph.HE},
       adsurl = {https://ui.adsabs.harvard.edu/abs/2022ApJ...927...61K}
}

@article{sahu2013,
author = {Sahu, Devendra and Anupama, G.C. and Chakradhari, N.},
year = {2013},
month = {05},
pages = {},
title = {One year of monitoring of the Type IIb supernova SN 2011dh},
volume = {433},
journal = {Monthly Notices of the Royal Astronomical Society},
doi = {10.1093/mnras/stt647}
}

@ARTICLE{Woosley_1994,
       author = {{Woosley}, S.~E. and {Eastman}, Ronald G. and {Weaver}, Thomas A. and {Pinto}, Philip A.},
        title = "{SN 1993J: A Type IIb Supernova}",
      journal = {\apj},
         year = 1994,
        month = jul,
       volume = {429},
        pages = {300},
          doi = {10.1086/174319},
       adsurl = {https://ui.adsabs.harvard.edu/abs/1994ApJ...429..300W}
}

@article{Chevalier2010,
doi = {10.1088/2041-8205/711/1/L40},
url = {https://doi.org/10.1088/2041-8205/711/1/L40},
year = {2010},
month = {feb},
publisher = {The American Astronomical Society},
volume = {711},
number = {1},
pages = {L40},
author = {Chevalier, Roger A. and Soderberg, Alicia M.},
title = {TYPE IIb SUPERNOVAE WITH COMPACT AND EXTENDED PROGENITORS},
journal = {The Astrophysical Journal Letters}
}

@article{Maund2015,
    author = {Maund, J. R. and Arcavi, I. and Ergon, M. and Eldridge, J. J. and Georgy, C. and Cenko, S. B. and Horesh, A. and Izzard, R. G. and Stancliffe, R. J.},
    title = {Did the progenitor of SN 2011dh have a binary companion?★},
    journal = {Monthly Notices of the Royal Astronomical Society},
    volume = {454},
    number = {3},
    pages = {2580-2585},
    year = {2015},
    month = {10},
    issn = {0035-8711},
    doi = {10.1093/mnras/stv2098},
    url = {https://doi.org/10.1093/mnras/stv2098},
    eprint = {https://academic.oup.com/mnras/article-pdf/454/3/2580/4030187/stv2098.pdf},
}

@article{Ryder2018,
doi = {10.3847/1538-4357/aaaf1e},
url = {https://doi.org/10.3847/1538-4357/aaaf1e},
year = {2018},
month = {mar},
publisher = {The American Astronomical Society},
volume = {856},
number = {1},
pages = {83},
author = {Ryder, Stuart D. and Dyk, Schuyler D. Van and Fox, Ori D. and Zapartas, Emmanouil and Mink, Selma E. de and Smith, Nathan and Brunsden, Emily and Bostroem, K. Azalee and Filippenko, Alexei V. and Shivvers, Isaac and Zheng, WeiKang},
title = {Ultraviolet Detection of the Binary Companion to the Type IIb SN 2001ig},
journal = {The Astrophysical Journal}
}

@ARTICLE{Fransson1989,
       author = {{Fransson}, Claes and {Chevalier}, Roger A.},
        title = "{Late Emission from Supernovae: A Window on Stellar Nucleosynthesis}",
      journal = {\apj},
         year = 1989,
        month = aug,
       volume = {343},
        pages = {323},
          doi = {10.1086/167707},
       adsurl = {https://ui.adsabs.harvard.edu/abs/1989ApJ...343..323F}
}

@ARTICLE{Hachinger2012,
       author = {{Hachinger}, S. and {Mazzali}, P.~A. and {Taubenberger}, S. and {Hillebrandt}, W. and {Nomoto}, K. and {Sauer}, D.~N.},
        title = "{How much H and He is 'hidden' in SNe Ib/c? - I. Low-mass objects}",
      journal = {\mnras},
         year = 2012,
        month = may,
       volume = {422},
       number = {1},
        pages = {70-88},
          doi = {10.1111/j.1365-2966.2012.20464.x},
archivePrefix = {arXiv},
       eprint = {1201.1506},
 primaryClass = {astro-ph.SR},
       adsurl = {https://ui.adsabs.harvard.edu/abs/2012MNRAS.422...70H}
}

@ARTICLE{Dessart2012,
       author = {{Dessart}, Luc and {Hillier}, D. John and {Li}, Chengdong and {Woosley}, Stan},
        title = "{On the nature of supernovae Ib and Ic}",
      journal = {\mnras},
         year = 2012,
        month = aug,
       volume = {424},
       number = {3},
        pages = {2139-2159},
          doi = {10.1111/j.1365-2966.2012.21374.x},
archivePrefix = {arXiv},
       eprint = {1205.5349},
 primaryClass = {astro-ph.SR},
       adsurl = {https://ui.adsabs.harvard.edu/abs/2012MNRAS.424.2139D}
}

@ARTICLE{Williamson2021,
       author = {{Williamson}, Marc and {Kerzendorf}, Wolfgang and {Modjaz}, Maryam},
        title = "{Modeling Type Ic Supernovae with TARDIS: Hidden Helium in SN 1994I?}",
      journal = {\apj},
         year = 2021,
        month = feb,
       volume = {908},
       number = {2},
          eid = {150},
        pages = {150},
          doi = {10.3847/1538-4357/abd244},
archivePrefix = {arXiv},
       eprint = {2010.10528},
 primaryClass = {astro-ph.HE},
       adsurl = {https://ui.adsabs.harvard.edu/abs/2021ApJ...908..150W}
}

\bibliographystyle{aasjournal}

\end{document}